\documentclass[journal, 10pt]{IEEEtran}
\IEEEoverridecommandlockouts

\usepackage{amssymb}
\usepackage[cmex10]{amsmath}
\usepackage{stfloats}
\usepackage{graphicx}
\usepackage{subfigure}
\usepackage{tabularx}
\usepackage{verbatim}
\usepackage{url}
\usepackage{bm}
\usepackage{cite}
\usepackage{booktabs}
\usepackage[
    colorlinks=true,
    linkcolor=black,
    citecolor=black,
    urlcolor=black	
]{hyperref}  
\usepackage{balance}
\usepackage{algorithm}
\usepackage{algorithmic}
\usepackage{datetime}
\usepackage{epstopdf}

\usepackage{color}
\definecolor{myc1}{rgb}{0,0,0}

\usepackage[font=small,skip=2pt]{caption}

\begin{document}
\title{Modeling and Performance Analysis for Fluid Antenna System Enabled UAV Near-Field Communications}

\author{Hao Jiang, \IEEEmembership{Senior Member, IEEE,} Wangqi Shi, Zhentian Zhang, \IEEEmembership{Member, IEEE,}, Xusheng Zhu, \IEEEmembership{Member, IEEE,} \\
            Kai-Kit Wong, \IEEEmembership{Fellow, IEEE,}
           and Hyundung Shin, \IEEEmembership{Fellow, IEEE}
\vspace{-9mm}

\thanks{This work is supported by NSFC projects (No. 62471238 and 62371197). }
\thanks{H. Jiang is with the School of Cyber Science and Engineering, Southeast University, Nanjing 210096, P. R. China (e-mail: jiang.hao@seu.edu.cn).}
\thanks{W. Shi, and Z. Zhang are with the National Mobile Communications Research Laboratory, Southeast University, Nanjing 210096, P. R. China. Z. Zhang is also with the Hong Kong Polytechnic University, Hong Kong SAR, China. (e-mail: {shiwangqi@seu.edu.cn, zhentianzhangzzt@gmail.com}).}
\thanks{Xusheng Zhu is with the Department of Electronic and Electrical Engineering, University College London, WC1E 6BT London, U.K. (e-mail: { xusheng.zhu@ucl.ac.uk}).}
\thanks{K. K. Wong is with the Department of Electronic and Electrical Engineering, University College London, WC1E 7JE London, United Kingdom, and also with the Department of Electronic Engineering, Kyung Hee University, Yongin-si, Gyeonggi-do 17104, Republic of Korea. (e-mail: kai-kit.wong@ucl.ac.uk).}
\thanks{H. Shin is  with the Department of Electronics and Information Convergence Engineering, Kyung Hee University, Yongin-si, Gyeonggi-do 17104, Republic of Korea (e-mail: hshin@khu.ac.kr).}
}

\maketitle

\begin{abstract}
Fluid antenna systems (FASs) offer a promising solution for unmanned aerial vehicle (UAV) air-to-ground (A2G) communications by enabling reconfigurable radiation characteristics. Addressing the limitations of traditional models in capturing the dynamic port configuration of FAS and the near-field nature of UAV communications, this paper proposes a dynamic port-reconfigurable near-field channel model for FAS-assisted UAV-to-mobile user (MU) links. Furthermore, we develop a FAS-adaptive subarray partition scheme utilizing a greedy strategy. By decomposing line-of-sight (LoS) and non-line-of-sight (NLoS) components and integrating UAV motion dynamics with FAS port activation states, the proposed model accurately characterizes the non-uniform spatial distribution of near-field channels. The subarray partition scheme dynamically groups active ports to satisfy near-field conditions while significantly reducing computational complexity, supported by a dynamic update algorithm that efficiently handles subarray adjustments during port switching.
To avoid low effective gain and deep-fading ports in dense FAS configurations, a channel gain-based selection strategy is employed to prioritize high-gain ports.
We derive and analyze the modeling accuracy and channel capacity, investigating the impact of FAS dimensions, port spacing, active port count, and UAV dynamics on system performance. Finally, the computational complexity of the subarray partition scheme is evaluated, verifying its advantages for real-time applications and providing a theoretical foundation for the design and analysis of FAS in dynamic scenarios.
\end{abstract}

\IEEEpeerreviewmaketitle

\vspace{-2mm}
\section{Introduction}

\subsection{Background and Related Works}

\IEEEPARstart{T}{he} core demands of the sixth generation (6G) wireless communication networks focus on ultra-high-speed data transmission, millisecond-level ultra-low latency, and ultra-reliable connections \cite{Saad-2020, Wang-2023}. Concurrently, 6G is required to seamlessly adapt to complex dynamic scenarios, such as space-air-ground integration and heterogeneous network fusion, providing robust communication guarantees for emerging applications like smart city emergency response, remote area coverage, and the industrial Internet of Things (IoT) \cite{Liu-2021, Aggarwal-2021}. In 6G air-to-ground (A2G) communication systems, unmanned aerial vehicles (UAVs) have emerged as pivotal carriers for critical tasks due to their flexible deployment and wide-area coverage capabilities. However, the high mobility of UAVs induces strong time-varying characteristics in communication channels. Moreover, during transmission, the line-of-sight (LoS) path is prone to obstruction, while the non-line-of-sight (NLoS) components exhibit complex, variable, and unpredictable behaviors \cite{Jiang-2025DL}. These challenges impose stringent requirements on the dynamic adaptability and spatial resource utilization efficiency of antenna systems.

\begin{figure}[!t]
	\centering\label{key}
	\includegraphics[width=\columnwidth]{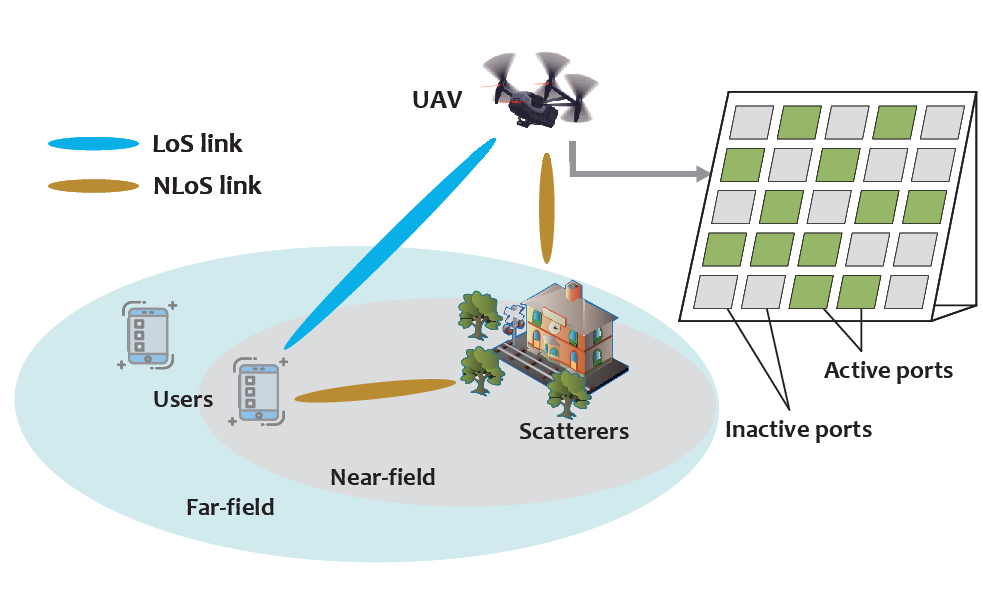}
	\caption{A schematic diagram of FAS-assisted UAV-to-MU wireless communication scenario.}\label{fig1}
	\vspace{-3mm}
\end{figure}

{\tiny }
Multiple-input multiple-output (MIMO) technology, a cornerstone of fifth-generation (5G) wireless communication, has played a crucial role in enhancing spectral efficiency and expanding channel capacity by leveraging spatial diversity gains \cite{Wang-2024MIMO, ZhangMIMO2,Wang-2024, ZhangMIMO}. However, when applied to UAV-assisted A2G scenarios, MIMO faces inherent limitations. First, traditional MIMO mandates a minimum antenna spacing of half a wavelength to mitigate mutual coupling, which is critical for maintaining signal independence and diversity gain \cite{Wong-2021}. Unfortunately, the limited payload and strict size constraints of UAVs make deploying a sufficient number of independent antenna elements impractical, directly restricting the scalability of MIMO spatial gains. Second, the number of radio-frequency (RF) chains in MIMO systems typically scales proportionally with the number of antenna elements. This expansion significantly increases hardware costs and power consumption, contradicting the lightweight and low-power design requirements of UAVs \cite{Jiang-2025FAS, Zhang-2025JSAC, Ghadi-2025}.

As an emerging wireless communication technology for 6G, the fluid antenna system (FAS) is distinguished by its flexible and reconfigurable radiation characteristics. This feature relaxes the strict fixed antenna spacing requirements of MIMO, enabling flexible adjustment of signal radiation patterns through dynamic activation and intelligent port reconfiguration \cite{Wong2020, Wong-2020fas, Zhang-2026,FAA2, FAA3, CKJ1,CKJ2,CKJ3,Zhang-2025,AD_z}. This not only accommodates the limited payload space of UAVs but also mitigates the high cost and power consumption issues associated with the rigid coupling between RF chains and antenna elements in MIMO \cite{Wong-2022, Alvim2023on, New2023fluid}. Consequently, FAS provides an innovative solution to overcome the bottlenecks of MIMO in UAV-assisted A2G communication. The authors in \cite{Zhang-2025CSI,Zhang-2025CSI2} proposed an algorithm based on geographical priors, enabling low-complexity and high-precision channel state information (CSI) acquisition in FAS, thereby leveraging high spatial gain in compact spaces. In \cite{Shen-2025}, the authors deployed FAS on a UAV, dynamically adjusting and jointly optimizing relevant parameters based on position to maximize the downlink rate. Furthermore, the authors in \cite{Abdou-2024} applied FAS to a UAV relay system, jointly optimizing active ports, power allocation, and UAV altitude. Their scheme maximizes the sum rate for ground users while meeting minimum rate requirements, outperforming traditional antenna schemes. Despite these advances, current research on FAS predominantly focuses on cellular network scenarios, and existing channel models fail to fully address the specific requirements of A2G communication. Therefore, developing a general FAS-assisted A2G channel model capable of adapting to various communication conditions remains an urgent priority.

Although FAS offers an effective pathway to overcome MIMO limitations in UAV-assisted A2G communication, its practical implementation is constrained by the challenges of near-field communication. Specifically, influenced by factors such as the UAV's antenna aperture, flight altitude, and coverage range, A2G communication between UAVs and mobile users (MUs) often undergoes dynamic switching between near- and far-field regions \cite{Han-2024}. The near-field region is defined by the Rayleigh distance $L = 2D^2/\lambda$, where $D$ and $\lambda$ represent the antenna aperture and wavelength, respectively. When the distance between the receiver and the UAV is less than $L$, the system operates in the near-field regime \cite{Cui-2024, Liu-2023, Ruan-2024}. The near-field environment differs fundamentally from the far-field: signal distribution in the spatial domain exhibits significant non-uniformity. This exacerbates the time-varying characteristics of the channel response and directly affects beamforming accuracy and signal diversity gain. Consequently, traditional channel modeling methods based on far-field assumptions are ill-suited for these scenarios, limiting overall communication performance.

Existing literature has extensively investigated near-field communication channel characteristics. Some studies have explored near-field propagation mechanisms for traditional arrays, such as uniform planar arrays (UPAs), and proposed corresponding channel modeling methods. For instance, the authors in \cite{Wan-2024} proposed a near-field non-stationary channel modeling scheme based on electromagnetic scattering theory and designed an estimation algorithm incorporating electromagnetic priors. In \cite{Yang-2024}, a large-scale MIMO multi-domain non-stationary channel model was constructed, jointly considering mutual coupling, antenna efficiency, and near-field directional vectors, with measurements verifying its statistical accuracy. Furthermore, \cite{Jiang-2025TWC, Jiang-2025IoT} proposed a subarray partition scheme by mechanically dividing the complete antenna array into multiple independent subarrays, each satisfying far-field conditions to reduce complexity. While the concept of subarrays is theoretically applicable to FAS, mechanical partition methods possess inherent structural limitations and cannot match the dynamic port configuration of FAS.

Motivated by the limitations of existing studies, such as the mismatch between current FAS channel models and dynamic near-/far-field switching, the inability of far-field models to capture non-uniform spatial signal distribution, and the incompatibility of mechanical subarray schemes with flexible FAS port activation, this paper constructs a general near-field channel model tailored for FAS-assisted UAV communication. Additionally, we propose an efficient subarray partition scheme adapted to the dynamic port configuration of FAS. The proposed model addresses the non-uniform distribution of near-field signals while meeting the real-time requirements of dynamic UAV scenarios through a lightweight design.
The proposed framework also has potential applications in mobile edge computing (MEC) and digital twin-enabled UAV networks. Reliable and adaptive UAV links can support delay-sensitive computation offloading in MEC-enabled aerial computing, while dynamic channel tracking can facilitate physical-virtual synchronization and resource allocation in digital twin systems \cite{Peng-2022, Chen-2026, Zhou-2026}.

\subsection{Main Contributions}

Current research on FAS exhibits significant gaps. Existing studies and channel models generally fail to adapt to the dynamic characteristics of UAV-assisted A2G communication, particularly the frequent switching between near-field and far-field regions and the time-varying nature of the channel. Furthermore, traditional far-field modeling methods cannot address the non-uniform spatial signal distribution in FAS near-field scenarios, and fixed mechanical subarray partition schemes designed for MIMO/RIS are incompatible with the dynamic port activation of FAS. To bridge these gaps, this paper focuses on FAS-assisted UAV-to-MU near-field communication. The main contributions are summarized as follows:

\begin{itemize}
	\item We propose a dynamic port-reconfigurable near-field channel model for FAS-assisted UAV-to-MU communications. This model decomposes LoS and NLoS propagation paths, characterizes the real-time activation status of ports via a port matrix, and integrates UAV motion dynamics to accurately characterize the non-uniform distribution of near-field channels.
	\item We design a subarray partition scheme based on a greedy strategy. Unlike fixed mechanical partitions in traditional MIMO/RIS, this scheme dynamically groups active ports under Rayleigh distance constraints, reducing computational complexity while satisfying near-field conditions. A dynamic update algorithm is also developed to handle subarray adjustments during new port activation, with computational complexity quantified and decomposed by stages.
	\item We propose a FAS multi-active port selection strategy based on maximum channel gain to avoid low effective gain and deep-fading ports in dense FAS configurations.
This strategy prioritizes high-gain ports and avoids those with deep fading, resolving transmission instability caused by random activation and adapting to dynamic channel changes.
	\item We derive analytical methods for the modeling accuracy and channel capacity of the proposed model. Modeling accuracy is quantified relative to UPA, and channel capacity is derived within a MIMO framework. We systematically explore the impact of FAS dimensions, port spacing, number of active ports, and UAV dynamics on these metrics, providing quantitative support for FAS parameter optimization in A2G scenarios.
\end{itemize}

The subsequent structure of this paper is organized as follows: Section II elaborates on the proposed 3D FAS-assisted UAV-to-MU channel model, including the detailed design of the greedy strategy-based subarray partition scheme and its dynamic update mechanism. Section III focuses on performance analysis, covering the optimal port selection strategy, quantitative evaluation of modeling accuracy relative to UPA, and derivation of theoretical channel capacity. Section IV presents numerical simulation results to verify the impacts of FAS configuration parameters, active port ratio, and UAV dynamic characteristics on system performance. Finally, Section V summarizes the key findings of this research and concludes the paper, with supplementary derivations provided in the Appendix.

\textbf{Notation}: In this work, lowercase letters (e.g., $x$), boldface lowercase letters (e.g., $\mathbf{x}$), and boldface uppercase letters (e.g., $\mathbf{X}$) denote scalars, vectors, and matrices, respectively. $\Vert \cdot \Vert$, $(\cdot)^*$, and $[\cdot]^\text{T}$ signify the Frobenius norm, complex conjugate, and transpose operations, respectively. Finally, $j = \sqrt{-1}$ represents the imaginary unit, and $\mathbb{E}[\cdot]$ denotes the expectation operation.

\section{Proposed System Model}

\begin{figure}[!t]
	\centering
	\includegraphics[width=\columnwidth]{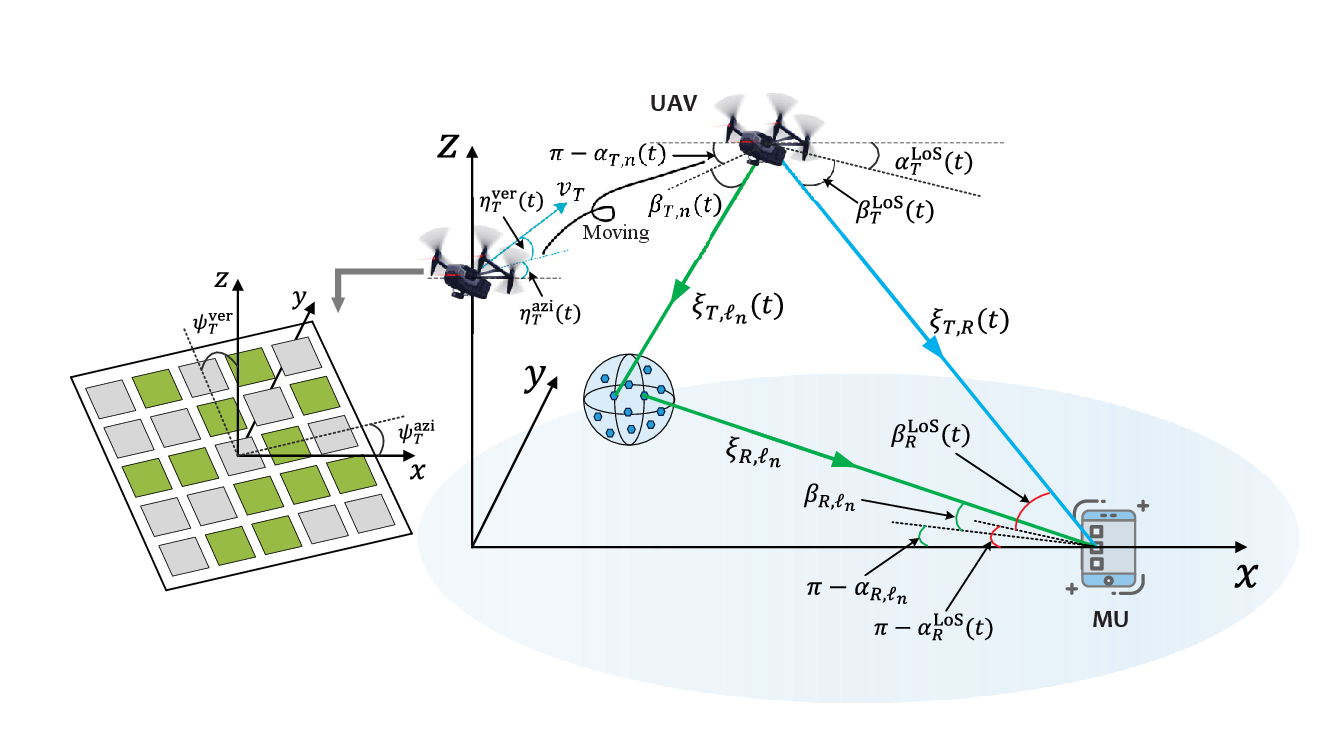}
	\caption{The 3D structural diagram of the proposed FAS-assisted UAV channel model.}\label{fig2}
	\vspace{-3mm}
\end{figure}

As shown in Figs.~\ref{fig1} and \ref{fig2}, we propose a three-dimensional (3D) FAS-assisted channel model for UAV-to-MU communications. 
The UAV transmitter is equipped with a planar FAS with physical dimensions $L_h \times L_v$ containing $P_h \times P_v$ ports, where $L_h=W_h \lambda$ and $L_v=W_v \lambda$, with $W_h$ and $W_v$ denoting the wavelength-normalized horizontal and vertical dimensions, respectively.
$\lambda$ denotes the carrier wavelength, and $P_h$ and $P_v$ denote the number of ports along the horizontal and vertical directions, respectively. The azimuth and elevation orientation angles of the FAS are defined as $\psi^{\text{azi}}_T$ and $\psi^{\text{ver}}_T$, respectively. The MU is equipped with $Q$ omni-directional uniform linear array (ULA) antennas, and the azimuth and elevation orientation angles of the antenna array are denoted by $\psi^{\text{azi}}_R$ and $\psi^{\text{ver}}_R$, respectively. Given the port dimensions $P_h$ and $P_v$, the spacing between adjacent ports in the FAS is given by
\begin{align}
	\Delta d_h = \frac{W_h\lambda}{P_h-1} ,
\end{align}
\begin{align}
	\Delta d_v = \frac{W_v\lambda}{P_v-1} .
\end{align}
Consequently, the initial position vector of the $(p_h, p_v)$-th port ($p_h=1,2,\dots,P_h, p_v=1,2,\dots,P_v$) relative to the center of the FAS can be expressed as
\begin{align}\label{eq:d_FAS}
	\mathbf{d}_{(p_h,p_v), T} =
	k_{p_h,p_v}
	\left [
	\begin{array}{ccc}
		\cos{\psi^{\text{ver}}_T} \cos{\psi^{\text{azi}}_T} \\ [0.125cm]
		\cos{\psi^{\text{ver}}_T} \sin{\psi^{\text{azi}}_T} \\ [0.125cm]
		\sin{\psi^{\text{ver}}_T}
	\end{array}
	\right ] \, ,
\end{align}
where $k_{p_h,p_v} = \sqrt{k^2_{p_h} \Delta d^2_h + k^2_{p_v} \Delta d^2_v}$ with $k_{p_h} = (P_h - 2p_h + 1)/2$ and $k_{p_v} = (P_v - 2p_v + 1)/2$. Furthermore, the proposed channel model considers $L$ clusters, where the $\ell$-th cluster ($\ell=1,2,\dots,L$) contains $\ell_N$ scatterers. The position of the center of the clusters is defined as $\mathbf{d}_{\text{cluster}} = [x_{\text{cluster}}, y_{\text{cluster}}, z_{\text{cluster}}]^{\text{T}}$.

In the FAS, ports are not all activated simultaneously; they operate in only two states: active or inactive. To characterize the real-time activation status of each port, we introduce a port matrix $\mathbf{S} (t)$, which shares the dimensions of the FAS port layout. Its elements are binary, where $1$ denotes an active port and $0$ denotes an inactive one, expressed as
\begin{align}
	\mathbf{S} (t) = [s_{p_h,p_v} (t)]_{P_v \times P_h},~~ s_{p_h,p_v} (t) \in \{0,1\} \, ,
\end{align}
where $s_{p_h,p_v} (t)$ indicates the status of the $(p_h,p_v)$-th port. This matrix facilitates a direct mapping between ports and their states. When combined with \eqref{eq:d_FAS}, it significantly simplifies the calculation of position parameters for active ports.

In this work, the MU is assumed to be a low-mobility or quasi-static ground terminal, whose displacement within one channel update interval is much smaller than the spatial variation scale of the considered UAV-to-MU channel. Conversely, the UAV moves with a speed $v_T$, with azimuth and elevation angles of motion denoted by $\eta^\text{azi}_T$ and $\eta^\text{ver}_T$, respectively. In the established 3D coordinate system, the projection of the UAV's initial position on the ground serves as the origin. The positive $x$-axis points towards the MU, the $z$-axis is vertically upward, and the $y$-axis is determined by the right-hand rule. Let $H_0$ and $D_0$ be the initial altitude of the UAV and the horizontal distance to the MU, respectively. The time-varying position of the UAV is expressed as
\begin{align}
	\mathbf{d}_T (t) =
	\left [
	\begin{array}{ccc}
		0 \\ [0.125cm]
		0 \\ [0.125cm]
		H_0
	\end{array}
	\right ] \, +
	v_T t  \left [
	\begin{array}{ccc}
		\cos{\eta^\text{ver}_T} \cos{\eta^\text{azi}_T} \\ [0.125cm]
		\cos{\eta^\text{ver}_T} \sin{\eta^\text{azi}_T} \\ [0.125cm]
		\sin{\eta^\text{ver}_T}
	\end{array}
	\right ] \, ,
\end{align}
where $t$ denotes the flight time, and other parameters are illustrated in Fig.~\ref{fig2}. At the MU side, $\mathbf{d}_R = [D_0, 0, 0]^T$, and the position of the $q$-th antenna ($q=1,2,\dots,Q$) is given by
\begin{align}
	\mathbf{d}_q  = \mathbf{d}_R +
	k_q \left [
	\begin{array}{ccc}
		\cos{\eta^\text{ver}_R} \cos{\eta^\text{azi}_R} \\ [0.125cm]
		\cos{\eta^\text{ver}_R} \sin{\eta^\text{azi}_R} \\ [0.125cm]
		\sin{\eta^\text{ver}_R}
	\end{array} \right ] \, ,
\end{align}
where $k_q = (Q - 2q +1)/2$. Therefore, the real-time position of the $(p_h,p_v)$-th element in FAS can be derive by
\begin{align}\label{eq:d_phpv}
&\mathbf{d}_{p_h,p_v} (t) = \mathbf{d}_T (t) - \mathbf{d}_{(p_h,p_v),T} \nonumber \\[0.125cm]
& \quad = \left [
	\begin{array}{ccc}
		v_T t \cos{\eta^\text{ver}_T} \cos{\eta^\text{azi}_T} - k_{p_h,p_v} \cos{\psi^{\text{ver}}_T} \cos{\psi^{\text{azi}}_T} \\ [0.125cm]
		v_T t \cos{\eta^\text{ver}_T} \sin{\eta^\text{azi}_T} - k_{p_h,p_v} \cos{\psi^{\text{ver}}_T} \sin{\psi^{\text{azi}}_T} \\ [0.125cm]
		v_T t \sin{\eta^\text{ver}_T} + H_0 - k_{p_h,p_v} \sin{\psi^{\text{ver}}_T}
	\end{array} \right ] \, .
\end{align}
Equation \eqref{eq:d_phpv} clearly presents the relationship between the position of the $(p_h, p_v)$-th element in the FAS and the motion time of the UAV, which facilitates the subsequent derivation and calculation process.

\subsection{Subarray partition scheme for FAS}

\begin{figure}[!t]
	\centering
	\includegraphics[width=\columnwidth]{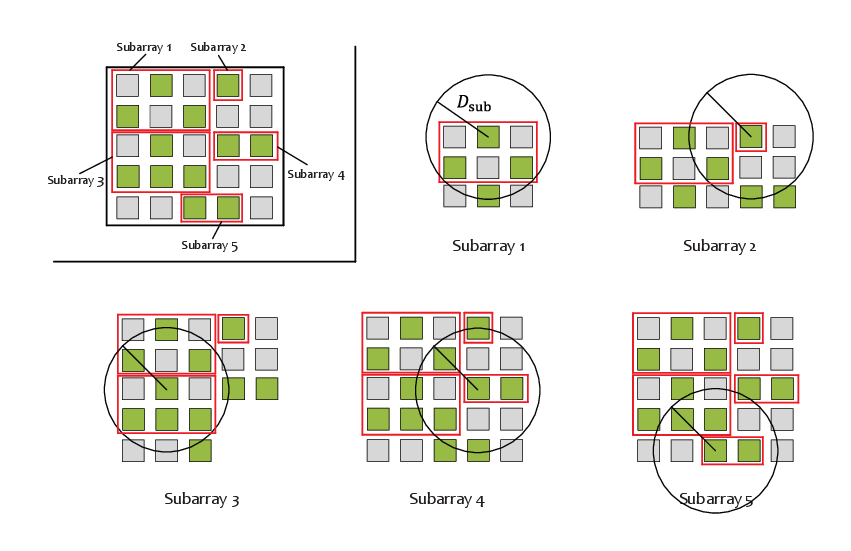}
	\caption{Illustration of the greedy strategy-based dynamic subarray partition scheme for a $5 \times 5$ planar FAS.}\label{subarray_partition}
	\vspace{-3mm}
\end{figure}

Unlike subarray partition schemes designed for MIMO or RIS, the core distinction of the FAS lies in its sparse port activation \cite{Espinosa2024, new2023information}. This characteristic precludes the use of fixed block division methods typical of MIMO or RIS arrays but affords the FAS greater flexibility in partition. This section provides a detailed explanation of the theoretical basis, implementation logic, and dynamic adaptability of the proposed scheme. The fundamental criterion for the partition is that, within any grouped subarray, the spatial distance between any two ports must not exceed the Rayleigh distance.

We first derive the maximum aperture required for a subarray to satisfy the far-field condition. Let $P^{\text{sub}}_{h,\text{max}}$ and $P^{\text{sub}}_{v,\text{max}}$ be the number of ports in the maximum subarray along the horizontal and vertical directions, respectively. The Rayleigh distance of this maximum subarray is
\begin{align}\label{eq:L_sub_max}
	L^{\text{sub}}_{\text{max}} = \frac{2 (D^{\text{sub}}_{\text{max}})^2}{\lambda} \, ,
\end{align}
\begin{align}
	D^{\text{sub}}_{\text{max}} = \sqrt{(P^{\text{sub}}_{h,\text{max}} - 1)^2 \Delta d^2_h + (P^{\text{sub}}_{v,\text{max}} - 1)^2 \Delta d^2_v} \, ,
\end{align}
where $D^{\text{sub}}_{\text{max}}$ denotes the maximum aperture. To ensure the receiver is located in the far-field region of all subarrays, $L^{\text{sub}}_{\text{max}}$ must be smaller than the distance from the subarray center to the receiver, denoted as $\xi_{\text{sub-max},R}$. Substituting \eqref{eq:L_sub_max} into this constraint yields
\begin{align}
	D^{\text{sub}}_{\text{max}} < \sqrt{\frac{\lambda \xi_{\text{sub-max},R}}{2}} \, .
\end{align}
This inequality demonstrates that if the maximum subarray aperture satisfies this condition, the far-field assumption holds for all valid subarrays.
It should be noted that the Rayleigh distance is used here as a standard and conservative boundary for identifying the near-field regime, rather than as an approximation of the actual propagation response. The partition threshold $D_{\text{sub}}$ controls the trade-off between modeling accuracy and computational complexity. A smaller $D_{\text{sub}}$ generates more subarrays and makes the proposed model closer to the spherical-wave model, but increases the computational burden. In contrast, a larger $D_{\text{sub}}$ reduces the number of subarray-wise calculations, while the local approximation error may increase. Therefore, $D_{\text{sub}}$ can be flexibly adjusted according to practical requirements under the Rayleigh distance-based subarray constraint.

Fig. \ref{subarray_partition} illustrates the proposed subarray partition scheme for a $5\times5$ FAS. The algorithm employs a greedy strategy: First, a partition aperture threshold $D_{\text{sub}} < D^{\text{sub}}_{\text{max}}$ is set. Starting with the first unpartitioned active port as the center, all adjacent active ports within a distance $D_{\text{sub}}$ are grouped into the same subarray. This process repeats, selecting a new starting port from the remaining unpartitioned set, until all active ports are assigned.

\begin{algorithm}[t!]
	\renewcommand{\algorithmicrequire}{\textbf{Input:}}
	\renewcommand{\algorithmicensure}{\textbf{Output:}}
	\caption{Subarray partition scheme}
	\label{subarray partition shceme}
	\begin{algorithmic}[1]
		\REQUIRE Partition aperture $D_{\text{sub}}$, port activation matrix $ \mathbf{S}(t) $;
		\ENSURE The set of subarrays $\text{G}_{\text{sub}}$, subarray center coordinates $\text{C}_{\text{center}}$;
		
		\STATE Initialize mark matrix $ \mathbf{M}_{\text{mark}} $ (same size as $ \mathbf{S}(t) $, all elements $ \text{False} $), $ \text{G}_{\text{sub}} = []$, $\text{C}_{\text{center}} = []$;
		\FOR{port $ (p_h, p_v) $ in $ \mathbf{S}(t) $}
		\IF{ $ \mathbf{S}(t)[p_h, p_v] == 1 $ and $ \mathbf{M}_{\text{mark}}[p_h, p_v] == \text{False} $ }
		\STATE $ \text{G}_{\text{sub}}^{\text{cur}} = [] $, $ \mathbf{M}_{\text{mark}}[p_h, p_v] = \text{True} $;
		\STATE Add port $ (p_h, p_v)$ to $\text{G}_{\text{sub}}^{\text{cur}}$ and $\text{C}_{\text{center}}$;
		
		\FOR{unmarked port $ (\gamma_h, \gamma_v) $ in $ \mathbf{S}(t) $}
		\IF{ $ \mathbf{S}(t)[\gamma_h, \gamma_v] == 1 $ and distance between $ (p_h, p_v) $ and $ (\gamma_h, \gamma_v) \leq D_{\text{sub}} $ }
		\STATE $ \mathbf{M}_{\text{mark}}[\gamma_h, \gamma_v] = \text{True} $;
		\STATE Add port $ (\gamma_h, \gamma_v) $ to $ \text{G}_{\text{sub}}^{\text{cur}} $;
		\ENDIF
		\ENDFOR
		
		\STATE Add $ \text{G}_{\text{sub}}^{\text{cur}} $ to $ \text{G}_{\text{sub}} $;
		\ENDIF
		\ENDFOR
	\end{algorithmic}
\end{algorithm}

The complete workflow is presented in Algorithm \ref{subarray partition shceme}. The algorithm outputs a structured collection of subarrays, $\text{G}_\text{sub}$, where each element represents an independent, non-overlapping subarray satisfying far-field conditions. We denote $\text{G}_\text{sub}$ as
\begin{align}\label{eq:G_sub}
	\text{G}_\text{sub} = \{ \text{G}^1_\text{sub}, \text{G}^2_\text{sub}, \dots, \text{G}^{\varepsilon}_\text{sub} \} \, ,
\end{align}
indicating that $\varepsilon$ subarrays are generated. A key feature is that all ports within a subarray share identical distance and angle parameters relative to the receiver. To simplify notation, we introduce representative parameters $\mathbf{d}_{\text{G}^1_\text{sub}}, \mathbf{d}_{\text{G}^2_\text{sub}}, \dots, \mathbf{d}_{\text{G}^{\varepsilon}_\text{sub}}$, which serve as the unified coordinates for each subarray. Given the UAV's motion, these coordinates are time-varying:
\begin{align}\label{eq:d_p_sub}
	\mathbf{d}_{\text{G}^{p_{\text{sub}}}_\text{sub}} (t) = \mathbf{d}_{\text{G}^{p_{\text{sub}}}_\text{sub}} + v_T t  \left [
	\begin{array}{ccc}
		\cos{\eta^\text{ver}_T} \cos{\eta^\text{azi}_T} \\ [0.125cm]
		\cos{\eta^\text{ver}_T} \sin{\eta^\text{azi}_T} \\ [0.125cm]
		\sin{\eta^\text{ver}_T}
	\end{array}
	\right ] \, ,
\end{align}
where $p_{\text{sub}} = 1,2,\dots,\varepsilon$ indexes the subarray.

This structured set $\text{G}_\text{sub}$ integrates the partition results efficiently. To determine the subarray membership of any active port, one simply traverses $\text{G}_\text{sub}$ to index the target subarray. Since active ports within a subarray share spatial parameters, the shared coordinates from \eqref{eq:d_p_sub} are used for channel gain calculations, avoiding redundant computations for individual ports.
Since Algorithm 1 adopts a greedy strategy, the partition result may depend on the scanning order of active ports. In this paper, a fixed row/column scanning order is adopted to ensure deterministic and reproducible partition results.

The computational complexity of Algorithm \ref{subarray partition shceme} comprises three parts:
\begin{itemize}
	\item  \textbf{Initialization:} Constructing $\mathbf{S} (t)$ and $\mathbf{M}_{\text{mark}}$ requires $O(P_hP_v)$.
	\item \textbf{Port Traversal:} Checking the ``active and unmarked'' status for all $P_h \times P_v$ ports takes $O(P_hP_v)$, as each check is $O(1)$.
	\item  \textbf{Subarray Partition:} For each unmarked active port, comparing distances with other active ports takes $O(P_{\text{act}}^2)$ in the worst case, where $P_{\text{act}}$ is the number of active ports.
\end{itemize}
Thus, the total complexity is
\begin{align}\label{eq:complexity}
	C_{\text{partition}} = O(P_hP_v + P_{\text{act}}^2) \, .
\end{align}

Since FAS dynamically adjusts active ports, subarray partitions must be updated upon new port activations. To minimize complexity, Algorithm \ref{subarray partition shceme} records the center port $\mathbf{C}_{\text{center}}$ of each subarray. When a new port is activated, the system calculates its distance to existing centers. If the distance is within $D_{\text{sub}}$, the port joins that subarray. Otherwise, it forms a new subarray. This dynamic update is detailed in Algorithm \ref{new_port_subarray}.

\begin{algorithm}[t!]
	\renewcommand{\algorithmicrequire}{\textbf{Input:}}
	\renewcommand{\algorithmicensure}{\textbf{Output:}}
	\caption{Dynamic subarray update for new active ports}
	\label{new_port_subarray}
	\begin{algorithmic}[1]
		\REQUIRE Partition aperture $D_{\text{sub}}$, new active port $(p^{\text{new}}_h, p^{\text{new}}_v)$, Subarray set $\text{G}_\text{sub}$, center port coordinates $\text{C}_\text{center}$;
		\ENSURE Subarray set $\text{G}_\text{sub}$, center port coordinates $\text{C}_\text{center}$;
		
		\FOR {each  $\varsigma \in \text{C}_\text{center}$}
		\IF {distance between $(p^{\text{new}}_h, p^{\text{new}}_v)$ and $\varsigma \leq D_{\text{sub}}$}
		\STATE Add $(p^{\text{new}}_h, p^{\text{new}}_v)$ to the subarray corresponding to $\varsigma$, update $\text{C}_\text{center}$;
		\STATE Break;
		\ENDIF
		\ENDFOR
		\IF {no $\varsigma$ satisfies distance condition}
		\STATE Add port $(p^{\text{new}}_h, p^{\text{new}}_v)$ to $\text{G}_\text{sub}$ as a new subarray, add to $\text{C}_\text{center}$;
		\ENDIF
	\end{algorithmic}
\end{algorithm}

For a single new port, the complexity of Algorithm \ref{new_port_subarray} is $C_{\text{update}} = O (\varepsilon)$ in the worst case, and typically lower in practice. In practice, with the optimization of terminating the traversal immediately after finding a matching subarray, the average case complexity is even lower.

To verify efficiency, we compare this with the traditional spherical wave scheme, which calculates coordinates, angles, and distances for each active port independently. Assuming a unit complexity $C_{\text{port}}$, the spherical wave scheme's total complexity is
\begin{align}\label{eq:spherical_complexity}
	C_{\text{spherical}} (t) = C_{\text{port}} P_{\text{act}}I(t) \, ,
\end{align}
where $I(t)$ is the number of updates over time $t$. Here, $C_{\text{port}}$ used only as a unit metric for algorithmic complexity analysis and does not represent the hardware cost of a physical port, RF chain, switching circuit, or control module.
The complexity of the proposed scheme is
\begin{align}\label{eq:subarray_complexity}
	C_{\text{subarray}} (t) = C_{\text{partition}} + ( C_{\text{update}} + \varepsilon P_{\text{port}})I(t) \, .
\end{align}
Although the proposed scheme has a fixed initialization overhead, its complexity becomes significantly lower than the spherical wave scheme over time because $\varepsilon \ll P_{\text{act}}$, effectively reducing redundant computations. This advantage is critical for real-time UAV communications.
To further clarify the role of the proposed greedy subarray partition, we note that a global exhaustive-search benchmark can be constructed by enumerating all feasible partitions satisfying the aperture constraint and selecting the one with the minimum modeling error. However, such exhaustive enumeration has combinatorial complexity and is therefore impractical for dynamic FAS-enabled UAV scenarios, which motivates the proposed greedy partition as a low-complexity accuracy-complexity tradeoff.

Notably, the proposed greedy-based partition is not a global optimum. However, pursuing a global optimum would entail prohibitive complexity, unsuitable for dynamic scenarios. The greedy approach represents a practical trade-off, ensuring modeling accuracy while meeting the real-time constraints of FAS-enabled UAV near-field communication. Furthermore, lightweight AI-based methods with offline training and efficient online inference are also promising alternatives.

Under high UAV mobility, faster channel variations may require more frequent port switching and dynamic updates. The structure of Algorithm \ref{new_port_subarray} avoids repeated global repartitioning, although its practical execution also depends on the port-switching latency and channel update interval of the FAS hardware. Recent electronically reconfigurable FAS prototypes have achieved microsecond-level reconfiguration \cite{Liu2025FAS}, suggesting that this overhead is unlikely to dominate in the considered low-mobility UAV scenario.

\subsection{Channel Modeling}

In the propagation link from UAV to MU, signal propagation exhibits significant path differentiation. Specifically, before reaching the receiver, electromagnetic signals transmitted by the FAS traverse two paths with distinct physical properties and propagation mechanisms: one is the direct LoS path, and the other is the NLoS path reflected by scattering clusters. Based on the path independence assumption, the channel characteristics of the entire communication link can be decomposed and modeled. and the complete channel matrix of the proposed model can be expressed as:
\begin{align}\label{eq:H}
	\mathbf{H}_{\text{FAS}}(t,\tau) = \mathbf{H}_{\text{FAS}}^{\text{LoS}} (t,\tau) + \mathbf{H}_{\text{FAS}}^{\text{NLoS}} (t,\tau)\, ,
\end{align}
where $\tau$ is the path delay, and $\mathbf{H}_{\text{FAS}}^{\text{LoS}}$ and $\mathbf{H}_{\text{FAS}}^{\text{NLoS}}$ represent the LoS and NLoS channel matrices, respectively. The dimension of $\mathbf{H}_{\text{FAS}}(t,\tau)$ is $Q\times P^{\text{act}}_h P^{\text{act}}_v$. Considering the hardware-induced mutual coupling under sub-half-wavelength port spacing, the channel matrix with coupling effective can be modeled as $\mathbf{H}_{\text{eff}} (t,\tau) = \mathbf{H}_{\text{FAS}} (t,\tau) \mathbf{C}_{\text{FAS}} (t,\tau)$, where $\mathbf{C}_{\text{FAS}} (t,\tau)$ denotes the mutual coupling matrix. Since this paper mainly focuses on near-field propagation modeling and port selection rather than electromagnetic coupling characterization, the numerical analysis adopts the uncoupled baseline with $\mathbf{C}_{\text{FAS}} (t,\tau) = \mathbf{I}$, under which $\mathbf{H}_{\text{eff}} (t,\tau)$ reduces to $\mathbf{H}_{\text{FAS}}(t,\tau)$.

Based on the channel matrix in \eqref{eq:H}, to further explore and quantitatively analyze the intrinsic propagation characteristics of the channel, we conducted detailed mathematical derivation for each element in the channel matrix $\mathbf{H}_{\text{FAS}} (t,\tau)$. These elements essentially correspond to the complex impulse responses (CIRs) between different transmit-receive antenna pairs. According to the proposed subarray partition scheme, active ports within the same subarray share distance and angle parameters. Leveraging this characteristic, the calculation process can be simplified. Specifically, for the active ports of the same subarray, it is only necessary to uniformly compute the CIR once, without the need for repeated calculations for each active port. Based on this, combined with the subarray set output by Algorithm \ref{subarray partition shceme}, the finally obtained CIR can be expressed as
\begin{align}\label{eq:h_pq}
	h^{\text{FAS}}_{p_{\text{sub}},q} (t,\tau) &= \sqrt{\frac{K}{K+1}} h^{\text{LoS}}_{p_{\text{sub}},q} (t) \delta \left( \tau - \tau^{\text{LoS}}(t) \right) \nonumber \\[0.125cm]
	&+ \sqrt{\frac{1}{K+1}} h^{\text{NLoS}}_{p_{\text{sub}},q} (t) \delta \left( \tau - \tau^{\text{NLoS}}(t) \right),
\end{align}
where $K$ is the Rician factor, which is used to indicate the proportion of LoS and NLoS components in the channel. $h^{\text{LoS}}_{p_{\text{sub}},q} (t)$ and $h^{\text{NLoS}}_{p_{\text{sub}},q} (t)$ are the CIRs for the LoS and NLoS links from the $p_{\text{sub}}$-th UAV subarray to the $q$-th MU antenna. $\tau^{\text{LoS}}(t)$ and $\tau^{\text{NLoS}}(t)$ are the corresponding path delays, we have
\begin{align}
	\tau^{\text{LoS}}(t) = \xi_{T,R}/c \, ,
\end{align}
\begin{align}
	\tau^{\text{NLoS}}(t) = (\xi_{T,\text{cluster}} + \xi_{R,\text{cluster}})/c \, ,
\end{align}
where $\xi_{T,R} (t) = \Vert \mathbf{d}_R - \mathbf{d}_T (t) \Vert$ is the distance between UAV and MU. $\xi_{T,\text{cluster}} = \Vert \mathbf{d}_{\text{cluster}} - \mathbf{d}_T (t) \Vert $ and $\xi_{R,\text{cluster}} = \Vert \mathbf{d}_{\text{cluster}} - \mathbf{d}_R \Vert $ are the distances from UAV and MU to the center of clusters, respectively.

For the LoS component, the calculation formula of CIR is mainly obtained through the comprehensive phase superposition of the propagation distance, the positions of transmitting and receiving antennas, and the motion state of the UAV. Thus the complex CIR from the $p_{\text{sub}}$-th element in FAS at the UAV side to the $q$-th element in ULA at the MU side can be expressed as
\begin{align}\label{eq:h_LoS}
	h^{\text{LoS}}_{p_{\text{sub}},q} (t) &= e^{-j \frac{2 \pi}{\lambda} \xi_{T,R} (t) } \nonumber \\[0.125cm]
	&\times e^{j \frac{2\pi}{\lambda} k_{p_{\text{sub}},h} \Delta d_h \cos \big(\alpha^{\text{LoS}}_{T}(t) - \psi^{\text{azi}}_T \big) \cos\beta^{\text{LoS}}_{T}(t) \cos{\psi^{\text{ver}}_T} } \nonumber \\[0.125cm]
	&\times e^{j \frac{2\pi}{\lambda} k_{p_{\text{sub}},v} \Delta d_v \sin\beta^{\text{LoS}}_{T}(t) \sin{\psi^{\text{ver}}_T} } \nonumber \\[0.125cm]
	&\times  e^{j \frac{2\pi}{\lambda} k_q \delta_R \cos\big(\alpha^{\text{LoS}}_{R}(t) - \psi^{\text{azi}}_R \big) \cos\beta^{\text{LoS}}_{R}(t) \cos \psi^{\text{ver}}_R } \nonumber \\[0.125cm]
	&\times  e^{j \frac{2\pi}{\lambda} k_q \delta_R \sin\beta^{\text{LoS}}_{R}(t) \sin \psi^{\text{ver}}_R } \nonumber \\[0.125cm]
	&\times  e^{j \frac{2\pi}{\lambda} v_T t \cos\big(\alpha^{\text{LoS}}_{T}(t) - \eta^{\text{azi}}_T \big) \cos\beta^{\text{LoS}}_{T}(t) \cos\eta^{\text{ver}}_T } \nonumber \\[0.125cm]
	&\times  e^{j \frac{2\pi}{\lambda} v_T t \sin\beta^{\text{LoS}}_{T}(t) \sin\eta^{\text{ver}}_T  }  \, ,
\end{align}
where $k_{p_{\text{sub}},h} \Delta d_h$ and $k_{p_{\text{sub}},v} \Delta d_v$ are the horizontal and vertical offsets of the subarray from the FAS center. The angles of departure (AoDs), $\alpha^\text{LoS}_T (t)$ and $\beta^\text{LoS}_T (t)$, are
\begin{align}\label{eq:alpha_los}
	\alpha^{\text{LoS}}_{T}(t) = \tan^{-1} \frac{d_{q,y} -  \mathbf{d}_{\text{G}^{p_{\text{sub}}}_\text{sub}, y} (t)}{d_{q,x} -  \mathbf{d}_{\text{G}^{p_{\text{sub}}}_\text{sub}, x} (t) } \, ,
\end{align}
\begin{align}\label{eq:beta_los}
	&\beta^{\text{LoS}}_{T}(t) = \nonumber \\[0.125cm]
	&\tan^{-1} \frac{d_{q,z} -  \mathbf{d}_{\text{G}^{p_{\text{sub}}}_\text{sub}, z} (t)}{ \sqrt{(\mathbf{d}_{\text{G}^{p_{\text{sub}}}_\text{sub}, x}(t) - d_{p,x}(t))^2 + (\mathbf{d}_{\text{G}^{p_{\text{sub}}}_\text{sub}, y} (t)- d_{p,y}(t))^2} } \, .
\end{align}
where $\mathbf{d}_{\text{G}^{p_{\text{sub}}}_\text{sub}, x} (t)$, $\mathbf{d}_{\text{G}^{p_{\text{sub}}}_\text{sub}, y} (t)$, and $\mathbf{d}_{\text{G}^{p_{\text{sub}}}_\text{sub}, z} (t)$ are the the components of vector $\mathbf{d}_{\text{G}^{p_{\text{sub}}}_\text{sub}} (t)$ along the $x$, $y$, and $z$-axis.
$\mathbf{d}_{p, x} (t)$, $\mathbf{d}_{p, y} (t)$, and $\mathbf{d}_{p, z} (t)$ are the the components of vector $\mathbf{d}_p (t)$ along the $x$, $y$, and $z$-axis.
Furthermore, $\alpha^\text{LoS}_R (t)$ and $\beta^\text{LoS}_R (t)$ are the time-varying angles of arrival of the transmitted waves in the azimuth and vertical planes, respectively. Based on the geometry properties, their relationship with the angles of departure can be expressed as $\alpha^\text{LoS}_R (t) = \pi - \alpha^\text{LoS}_T (t)$ and $\beta^\text{LoS}_R (t) = \beta^\text{LoS}_T (t)$, respectively. Although a planar-wave approximation is adopted within each subarray in (20), the overall near-field characteristics are maintained through the channel parameters such as the propagation distance and arrival/departure angles related to each subarray, which are calculated based on the center of each subarray.
%

For the NLoS component, the complex CIR from the $p_{\text{sub}}$-th subarray in FAS at the UAV side to the $q$-th element in ULA at the MU side can be expressed as
\allowdisplaybreaks[1] 
\begin{align}\label{eq:h_NLoS}
	h^{\text{NLoS}}_{p_{\text{sub}},q} (t) &= \sum_{\ell \in L} \sum^{\ell_N}_{n = 1} e^{j \varphi_{\ell_n} -j \frac{2 \pi}{\lambda} \big(\xi_{T,\ell_n} (t) + \xi_{R,\ell_n} \big) } \nonumber \\[0.125cm]
	&\times e^{j \frac{2\pi}{\lambda} k_{p_{\text{sub}},h} \Delta d_h \cos\big(\alpha^{\text{NLoS}}_{T,\ell_n}(t) - \psi^{\text{azi}}_T \big) \cos\beta^{\text{NLoS}}_{T,\ell_n}(t)  \cos{\psi^{\text{ver}}_T} }  \nonumber \\[0.125cm]
	&\times e^{j \frac{2\pi}{\lambda} k_{p_{\text{sub}},v} \Delta d_v \sin\beta^{\text{NLoS}}_{T,\ell_n}(t) \sin{\psi^{\text{azi}}_T} } \nonumber \\[0.125cm]
	&\times  e^{j \frac{2\pi}{\lambda} k_q \delta_R \cos\big(\alpha^{\text{NLoS}}_{R,\ell_n} - \psi^{\text{azi}}_R \big)\cos\beta^{\text{NLoS}}_{R,\ell_n} \cos \psi^{\text{ver}}_R} \nonumber \\[0.125cm]
	&\times  e^{j \frac{2\pi}{\lambda} k_q \delta_R \sin\beta^{\text{NLoS}}_{R,\ell_n} \sin \psi^{\text{ver}}_R } \nonumber \\[0.125cm]
	&\times  e^{j \frac{2\pi}{\lambda} v_T t \cos\big(\alpha^{\text{NLoS}}_{T,\ell_n}(t) - \eta^{\text{azi}}_T\big) \cos\beta^{\text{NLoS}}_{T,\ell_n} (t) \cos\eta^{\text{ver}}_T } \nonumber \\[0.125cm]
	&\times  e^{j \frac{2\pi}{\lambda} v_T t \sin\beta^{\text{NLoS}}_{T,\ell_n}(t) \sin\eta^{\text{ver}}_T  }  \, ,
\end{align}
where $\varphi_{\ell_n} \sim \text{U}[-\pi,\pi)$ is the random phase. $\xi_{T,\ell_n} (t)$ and $\xi_{R,\ell_n}$ are distances to the $n$-th scatterer in the $\ell$-th cluster:
\begin{align}
	\xi_{T,\ell_n} (t) &= \Vert \mathbf{d}_{\ell_n} - \mathbf{d}_{T} (t) \Vert,\\[0.125cm]
	\xi_{R,\ell_n} &= \Vert \mathbf{d}_{\ell_n} - \mathbf{d}_{R} \Vert,
\end{align}
where $\mathbf{d}_{\ell_n}$ is the scatterer position. The AoDs $\alpha_{T,\ell_n} (t)$ and $\beta_{T,\ell_n} (t)$ are given by
\begin{align}\label{eq:alpha_nlos}
	\alpha^{\text{NLoS}}_{T,\ell_n} (t) = \tan^{-1} \frac{y_{\ell_n} - \mathbf{d}_{\text{G}^{p_{\text{sub}}}_\text{sub}, y} (t)}{x_{\ell_n} -  \mathbf{d}_{\text{G}^{p_{\text{sub}}}_\text{sub}, x} (t) } \, ,
\end{align}
\begin{align}\label{eq:beta_nlos}
	&\beta^{\text{NLoS}}_{T,\ell_n}(t) = \nonumber \\[0.125cm]
	&\tan^{-1} \frac{z_{\ell_n} -  \mathbf{d}_{\text{G}^{p_{\text{sub}}}_\text{sub}, z} (t)}{ \sqrt{(\mathbf{d}_{\text{G}^{p_{\text{sub}}}_\text{sub}, x}(t) - x_{\ell_n})^2 + (\mathbf{d}_{\text{G}^{p_{\text{sub}}}_\text{sub}, y} (t)- y_{\ell_n})^2} } \, .
\end{align}

The AoAs $\alpha_{R,\ell_n}$ and $\beta_{R,\ell_n}$ are
\begin{align}\label{eq:alpha_r}
	\alpha^{\text{NLoS}}_{R,\ell_n} = \tan^{-1} \frac{y_{\ell_n} - \mathbf{d}_{q,y}}{x_{\ell_n} - \mathbf{d}_{q,x} } \, ,
\end{align}
\begin{align}\label{eq:beta_r}
	\beta^{\text{NLoS}}_{R,\ell_n} = \tan^{-1} \frac{z_{\ell_n} - \mathbf{d}_{q,z}}{ \sqrt{(\mathbf{d}_{q,x} - x_{\ell_n})^2 + (\mathbf{d}_{q,y} - y_{\ell_n})^2} } \, .
\end{align}

In this paper, we introduce von Mises distribution to determine the elevation and azimuth distributions of each scatter, which can be found from
\begin{align}
f( \alpha ) = \frac{e^{\kappa \cos{(\alpha - \mu_{\alpha})}}}{2\pi I_0(\kappa) } \, , \\[0.125cm]
f( \beta ) = \frac{e^{\kappa \cos{(\beta - \mu_{\beta})}}}{2\pi I_0(\kappa) } ,
\end{align}
where $\kappa$ is the environment factor, and a higher value of $\kappa$ indicates a more concentrated distribution of scatterers. $\mu_{\alpha}$ and $\mu_{\beta}$ are the mean values of azimuth and elevation angles, respectively. $I_0(\cdot)$ is the zero-order modified Bessel function.
Specifically, leveraging the von Mises distribution, initial arrival elevation and azimuth angle samples are generated to specify the angular direction of the scatterers. With scenario-specific parameters, initial scatterer-receiver distance samples are set. Finally, spatial coordinate transformation fuses these parameters to generate the scatterers’ initial spatial positions.

\vspace{-1mm}

\section{Performance Analysis of the Proposed Framework}

This section details the optimal port selection strategy and evaluates the performance of the proposed FAS channel model. Specifically, we analyze the modeling accuracy in comparison to traditional MIMO systems and derive the theoretical channel capacity.

\subsection{Optimal Port Selection}

The fundamental distinction between FAS and MIMO lies in the port spacing of FAS, which is typically less than half a wavelength. This sub-half-wavelength proximity induces mutual coupling between adjacent ports, resulting in a non-uniform spatial distribution of the channel. Consequently, while some ports reside in high-gain regions, others may suffer from deep fading, directly compromising transmission reliability \cite{Zhang-2025opt, New-2025opt, Xu2024}. Existing literature on FAS port selection predominantly follows the maximum channel gain criterion, aiming to guarantee transmission performance by locking onto high-gain ports. While this criterion is widely validated for single active port scenarios, we extend it to the multi-active port requirement of this study. Specifically, we select multiple ports with the highest channel gains as active ports to balance transmission performance and computational complexity. To facilitate this, we first define the channel matrix encompassing all ports as:
\begin{align}\label{eq:H_all_ports}
	\mathbf{H} (t,\tau) = \mathbf{H}^{\text{LoS}} (t,\tau) + \mathbf{H}^{\text{NLoS}} (t,\tau) \, ,
\end{align}
where $\mathbf{H}^{\text{LoS}} (t,\tau)$ and $\mathbf{H}^{\text{NLoS}} (t,\tau)$ denote the channel matrices for the LoS and NLoS propagation links, respectively. Notably, the channel matrix defined here differs from \eqref{eq:H}. Specifically, \eqref{eq:H_all_ports} is employed to determine the active ports, which subsequently form the basis for calculating the effective channel matrix in \eqref{eq:H}. The detailed derivation of \eqref{eq:H_all_ports} is provided in the APPENDIX.
It should be clarified that the proposed port selection strategy does not physically eliminate mutual coupling among adjacent ports. Instead, it prioritizes ports with high effective channel gains and avoids low-gain or deep-fading ports. It is also worth mentioning that inactive ports can be temporarily activated during the channel training stage through port switching. In addition, the CSI of inactive ports may also be inferred by exploiting the spatial correlation among densely distributed FAS ports. Based on the acquired or inferred CSI of candidate ports, the full channel matrix in (32) can then be constructed for optimal port selection.

\begin{figure}[!t]
	\centering
	\includegraphics[width=\columnwidth]{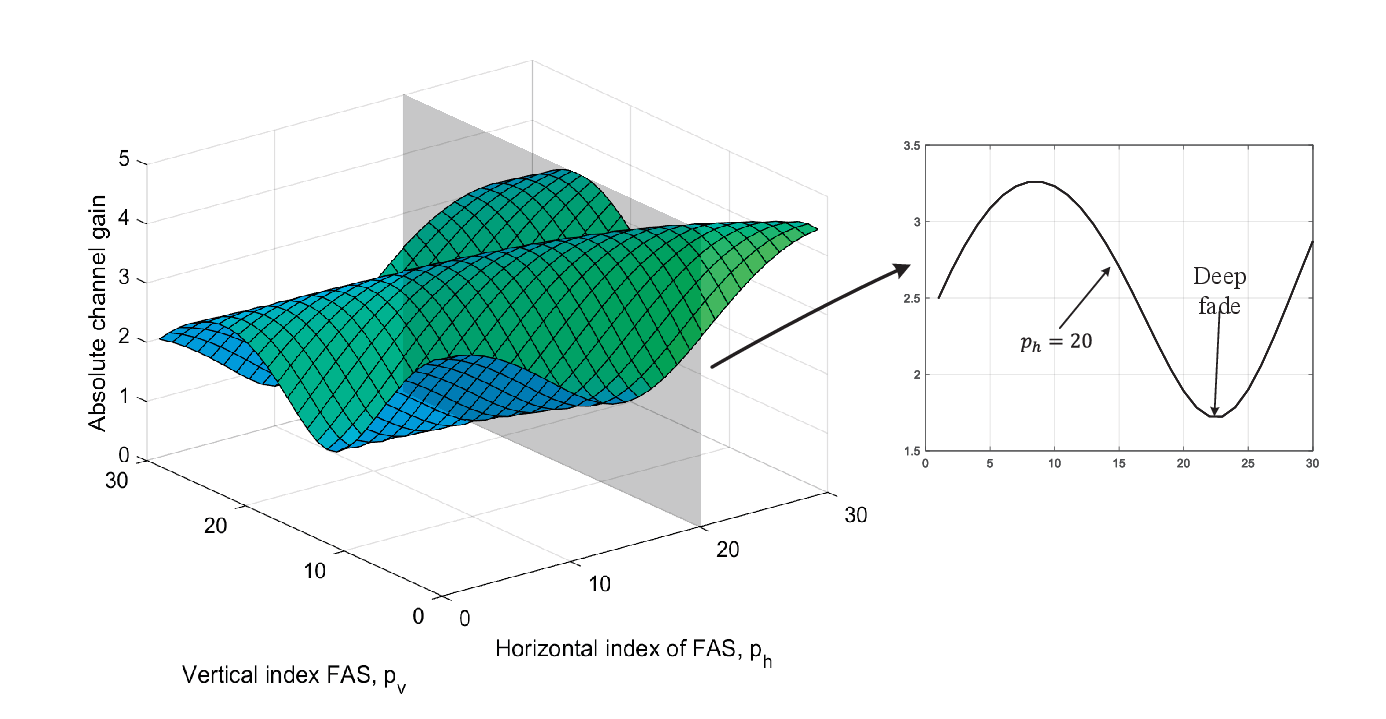}
	\caption{Spatial distribution of absolute channel gain for the $30 \times 30$ planar FAS}\label{channel_gain}
	\vspace{-3mm}
\end{figure}

Based on \eqref{eq:H_all_ports}, Fig. \ref{channel_gain} illustrates the channel gain of a $30 \times 30$ FAS, where the vertical axis represents the absolute channel gain. It is evident that significant variations in channel gain exist among antenna ports with different spatial parameters, reflecting the spatial non-uniformity of the channel. Observing the cross-section at fixed $p_h = 20$, the deep fading phenomenon is clearly visible. When the channel undergoes deep fading, the gain drops precipitously, leading to a substantial reduction in received signal power and a sharp increase in bit error rate, thereby severely degrading communication performance. Therefore, in FAS-based near-field communication, ``optimal port selection'' is a critical strategy for performance enhancement. By prioritizing high-gain ports and avoiding those in deep fading regions, the system ensures more stable and robust signal transmission.

\begin{figure}[!t]
	\centering
	\includegraphics[width=\columnwidth]{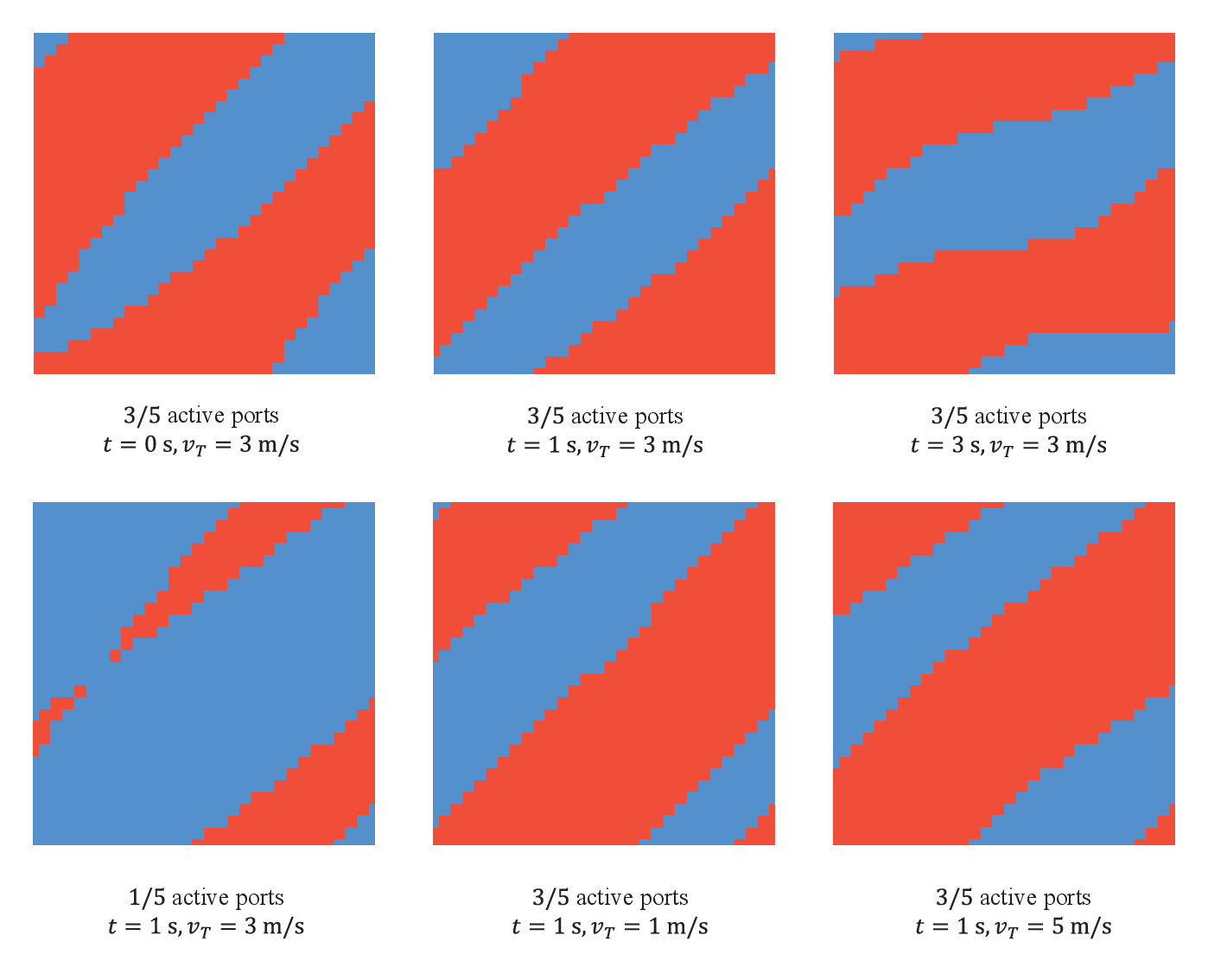}
	\caption{Dynamic spatial distribution of optimally selected active ports (red) and inactive ports (blue) for the $30 \times 30$ planar FAS.}\label{active_ports}
	\vspace{-3mm}
\end{figure}

As shown in Fig. \ref{active_ports}, the spatial distribution of active ports exhibits dynamic changes corresponding to the UAV's motion, where red indicates active ports and blue indicates inactive ones. Since the ``optimal port selection'' strategy is driven by channel gain, active ports are not randomly distributed but cluster in spatial regions exhibiting high channel gain. This clustering characteristic not only enables the FAS to actively bypass deep fading regions but also effectively minimizes the need for frequent global subarray re-partitioning, thereby meeting the real-time requirements of dynamic UAV communication scenarios. Additionally, the dynamic shift of the active port clusters in Fig. \ref{active_ports} over time reflects the adaptive reconfiguration capability of the FAS. Specifically, as the UAV moves, the time-varying propagation distances and dynamic angles of departure/arrival cause a continuous shift in the spatial position of high-gain regions. The FAS adapts by dynamically adjusting the active port set to track these high-gain regions continuously.

Equal power allocation is adopted among the selected active ports for the capacity analysis. Nevertheless, unequal channel gains may persist under severe near-field fading, making adaptive power allocation potentially more effective. Since this work primarily focuses on near-field channel modeling and high-gain port selection, equal power allocation is adopted to facilitate the analysis of these aspects.

\subsection{Channel Modeling Accuracy}

In the design and evaluation of fluid antenna wireless communication systems, quantifying the modeling deviation between FAS and traditional MIMO systems is crucial for validating the consistency between theoretical analysis and practical deployment.
Here, since the proposed FAS has a planar geometry, the corresponding uniform planar array (UPA)-based channel model is adopted as the reference benchmark, while all other conditions remain unchanged. Moreover, the same subarray processing procedure is applied to both models. Therefore, the UPA-based channel model provides a well-established reference for evaluating the proposed planar FAS channel model under identical communication conditions.
The normalized absolute error $\Delta$ is defined as:
\begin{align}\label{eq:Delta}
	\Delta = 10 \log_{10} \left\{\sum^{\varepsilon}_{p_{\text{sub}} = 1} \sum^{Q}_{q = 1} \frac{\vert h^{\text{FAS}}_{p_{\text{sub}},q} (t,\tau) - h^{\text{UPA}}_{p_{\text{sub}},q} (t, \tau) \vert}{\vert h^{\text{UPA}}_{p_{\text{sub}},q} (t, \tau) \vert} \right\},
\end{align}
where $h^{\text{UPA}}_{p_{\text{sub}},q} (t, \tau)$ represents the complex CIR of the baseline model. To ensure a fair comparison, the proposed subarray partition scheme is also applied to the baseline model. Eq. \eqref{eq:Delta} calculates the aggregate modeling error by traversing all subarrays. A smaller value of $\Delta$ indicates a higher degree of consistency between the channel characteristics of the proposed planar FAS and the UPA. Consequently, physical parameters of the FAS, such as port spacing, dimensions, and the number of active ports, significantly influence the modeling error. The impact of these parameters on error performance is analyzed in detail in the subsequent section. It is worth noting that the complexity-reduction capability of subarray-based modeling has been verified in our previous MIMO/RIS near-field studies, while the focus of this work is to adapt this idea to the dynamic sparse port activation mechanism of FAS-enabled UAV communications.

\subsection{Channel Capacity}

Channel capacity serves as another critical metric for measuring the theoretical performance limits of the proposed FAS. Given that the operating principle of the proposed planar FAS shares similarities with MIMO systems, the channel capacity can be formulated using the standard MIMO capacity expression:
\begin{equation}\label{eq:CP}
	C = \log_2 \left( \det\big(\mathbf{I}_{Q} + \frac{\rho_{\text{SNR}}}{P_{h,\text{act}}P_{v,\text{act}}} \overline{\mathbf{H}}_{\text{FAS}} (t,\tau) \overline{\mathbf{H}}^{\mathrm{H}}_{\text{FAS}} (t, \tau) \big)\right),
\end{equation}
where $\mathbf{I}_{Q}$ is the $Q \times Q$ identity matrix, and $\rho_{\text{SNR}}$ denotes the signal-to-noise ratio (SNR). $\overline{\mathbf{H}}_{\text{FAS}} (t,\tau)$ represents the normalized channel matrix, expressed as:
\begin{equation}\label{eq:CP_nor}
	\overline{\mathbf{H}}_{\text{FAS}} (t,\tau) = \mathbf{H}_{\text{FAS}} (t,\tau) \times \left\{ \frac{1}{P_{h,\text{act}}P_{v,\text{act}}Q} \Vert \mathbf{H}_{\text{FAS}} (t,\tau) \Vert_\text{F} \right\}^{-\frac{1}{2}},
\end{equation}
where $\Vert \cdot \Vert_\text{F}$ denotes the Frobenius norm. $P_{h,\text{act}}$ and $P_{v,\text{act}}$ are the number of active ports along the horizontal and vertical directions. Equations \eqref{eq:CP} and \eqref{eq:CP_nor} indicate that the channel capacity of the FAS is determined by the number of ports at both the transmitter and receiver. Furthermore, parameters such as port spacing, the number of active ports, and the proposed subarray partition scheme influence the capacity by modifying the structure of the channel matrix. The specific mechanisms of these influences are discussed in the following section.

\vspace{-2mm}

\section{Results and Discussion}

In this section, we present the numerical results regarding the modeling accuracy and channel capacity of the FAS system. The simulation parameters are configured as follows: $f_c = 5$ GHz, $H_0 = 20$ m, $D_0 = 60$ m, $Q = 4$, $\delta_R = \lambda/2$, $\psi^{\text{azi}}_R = \pi/2$, $\psi^{\text{ver}}_R = \pi/3$, $v_T = 3$ m/s, $\eta^{\text{azi}}_T = \eta^{\text{ver}}_T = \pi/3$, and $t = 2$ s. For the FAS, $P_h = P_v = 50$, $\psi^{\text{azi}}_T = \pi/2$, $\psi^{\text{ver}}_R = \pi/3$, $\Delta d_h = \Delta d_v = 2/5 \lambda $, and $3/5$ active ports. It is worth mentioning that the $50 \times 50$ FAS configuration is adopted as a large numerical benchmark to ensure that the considered setup clearly satisfies the near-field condition and to make the spherical-wave and spatially non-uniform channel characteristics more observable. In practical deployments, the FAS aperture and port number can be scaled according to the UAV payload, platform size, and carrier frequency.

The environmental parameter $\kappa$ characterizes the angular concentration of scatterers, with smaller values representing broader angular spreads and richer scattering. In practical deployments, $\kappa$ can be selected according to standardized or measurement-based angular spread characteristics of the considered propagation environment.

\subsection{Channel Modeling Accuracy}

\begin{figure}[!t]
	\centering
	\includegraphics[width=\columnwidth]{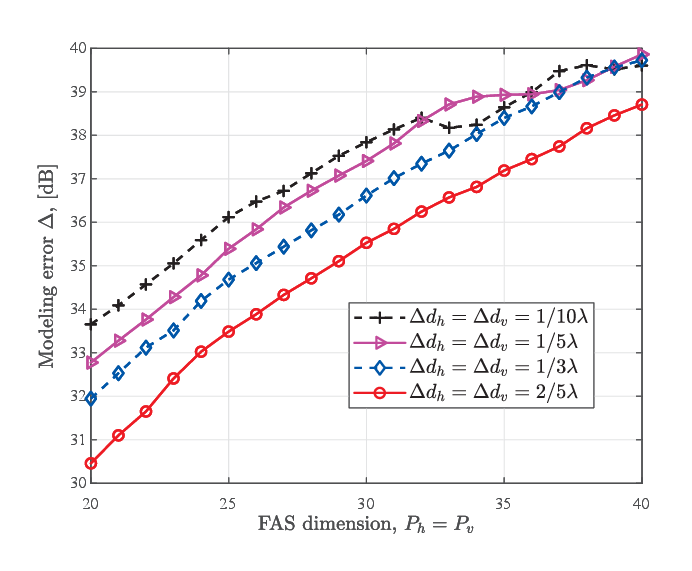}
	\caption{Modeling error performance of the proposed FAS-assisted UAV-to-MU channel model against the port spacing.}\label{fig3}
	\vspace{-3mm}
\end{figure}

Based on the modeling accuracy derivation provided in the previous section, Fig. \ref{fig3} illustrates the variation trends of the modeling error with respect to FAS dimensions and investigates the impact of different port spacings. It is observed that for any given port configuration, the modeling error increases as the FAS dimensions expand. Conversely, for a fixed FAS dimension, a smaller port spacing results in a larger modeling error. Furthermore, when the FAS dimension exceeds $30\times30$, the randomness and complexity of channel fading are intensified. Concurrently, excessively small port spacing in large-scale FAS induces stronger mutual coupling effects between ports, which in turn causes the fluctuations observed in the two curves with the smallest port spacings.

\begin{figure}[!t]
	\centering
	\includegraphics[width=\columnwidth]{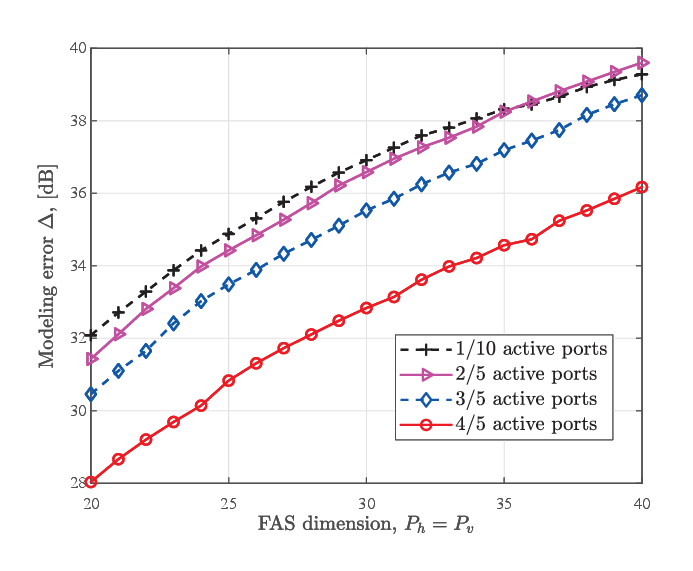}
	\caption{Modeling error performance of the proposed FAS-assisted UAV-to-MU channel model against the number of active ports.}\label{fig4}
	\vspace{-3mm}
\end{figure}

Fig. \ref{fig4} depicts the modeling accuracy of the proposed FAS-enabled channel model as a function of the number of active ports. It is evident that an increase in the number of active ports leads to a reduction in modeling errors. This phenomenon is closely related to the proposed subarray partition scheme and the optimal port selection strategy. Specifically, as reflected in the channel gain distribution (Fig. \ref{channel_gain}), ports with high channel gain exhibit a spatially concentrated pattern. When the number of active ports is small, the limited selection range results in a relatively concentrated distribution of partitioned subarrays, making it difficult to comprehensively and precisely capture the global channel characteristics. However, as the number of active ports increases, the partitioned subarrays cover a wider spatial area. This not only incorporates more key ports with high channel gain, thereby enhancing the acquisition of effective channel information, but also ensures a more complete match with the spatial correlation and distribution characteristics of the channel. Nevertheless, in practical system design, the selection of the number of active ports requires a trade-off between modeling accuracy and system performance to avoid potential deep fading issues.

\begin{figure}[!t]
	\centering
	\includegraphics[width=\columnwidth]{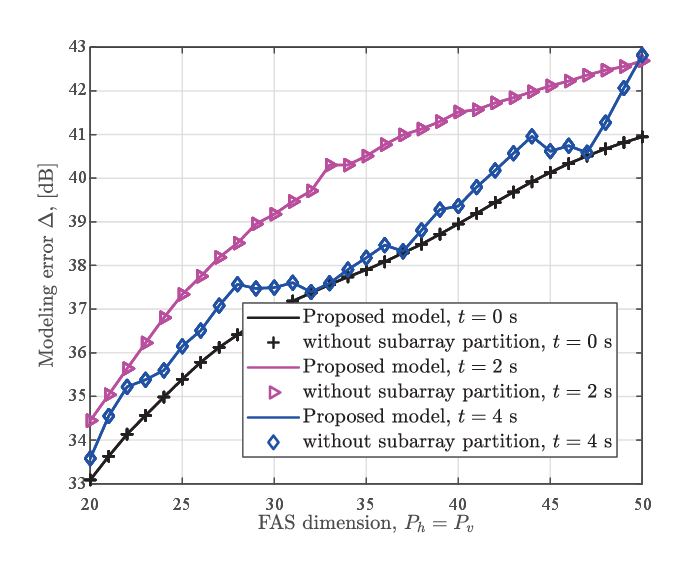}
	\caption{Time-varying modeling error performance of the proposed subarray partition scheme against FAS the dimensions} \label{modeling_accuracy_subarray}
	\vspace{-3mm}
\end{figure}

Fig. \ref{modeling_accuracy_subarray} evaluates the performance of the proposed FAS-enabled channel model based on the subarray partition scheme. From the time-domain perspective, the modeling error curve varies with the UAV's motion time, fully reflecting the non-stationarity of the channel. This non-stationarity originates from the interaction between the dynamic spatial position changes induced by UAV motion and the propagation environment. Furthermore, regardless of the increase in FAS size or the progression of motion time, the modeling accuracy of the proposed subarray partition scheme closely aligns with that of the scheme without subarray partition, exhibiting a similar error growth rate. This further demonstrates that the proposed subarray partition scheme not only maintains stable modeling accuracy but also effectively adapts to the temporal non-stationary characteristics of dynamic channels, making it highly suitable for practical applications in dynamic scenarios such as UAV communications.

The total number of generated subarrays $\epsilon$ may also depend on the scanning order in Algorithm 1, since different scanning patterns can lead to different center-port selections and grouping results, thereby affecting the computational load. In this work, a fixed row-by-column scanning order is adopted for deterministic and reproducible partitioning.

\subsection{Channel Capacity}

\begin{figure}[!t]
	\centering
	\includegraphics[width=\columnwidth]{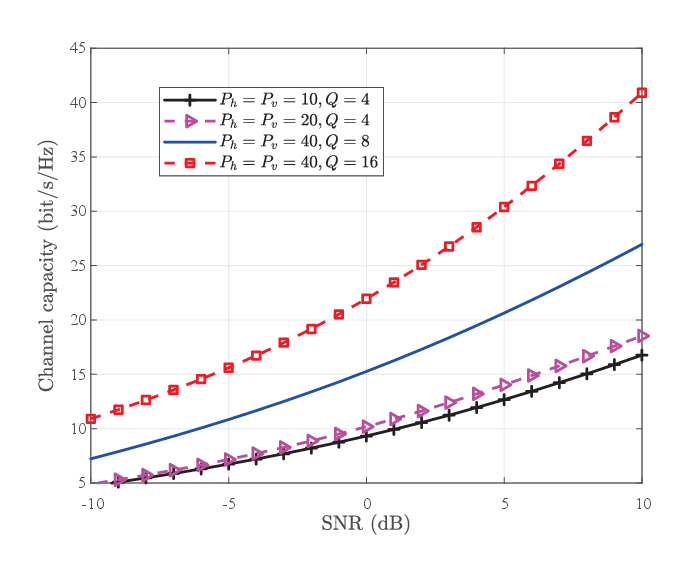}
	\caption{Channel capacity of the proposed FAS-assisted UAV-to-MU channel model with different antenna dimensions.}\label{channel_capacity_antenna_dimension}
	\vspace{-3mm}
\end{figure}

Utilizing \eqref{eq:CP} and \eqref{eq:CP_nor}, Fig. \ref{channel_capacity_antenna_dimension} presents the channel capacity of the proposed FAS-enabled channel model with respect to different antenna dimensions at the UAV and MU sides. It is evident that the channel capacity is directly proportional to the SNR, which verifies the validity of the proposed channel model. Another significant finding is that the growth rate of channel capacity is more sensitive to the number of receiving antennas. Specifically, when the number of receiving antennas increases from $8$ to $16$, the channel capacity increases by approximately 50\%. To achieve a comparable capacity gain by adjusting the transmitter, the number of transmitting antennas would need to be increased several times over. This phenomenon is explained by the fundamental property that MIMO channel capacity scales linearly with $\min\{P,Q\}$, where $P$ and $Q$ denote the number of antennas at the transmitter and receiver sides, respectively. These observations further corroborate the accuracy of the proposed FAS-enabled channel model.

\begin{figure}[!t]
	\centering
	\includegraphics[width=\columnwidth]{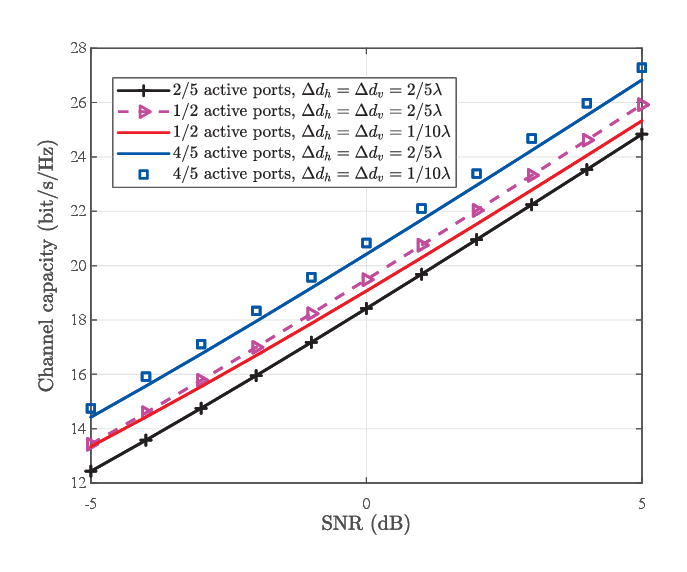}
	\caption{Channel capacity of the proposed FAS-assisted UAV-to-MU channel model with different active ports.}\label{channel_capactiy_active_ports}
	\vspace{-3mm}
\end{figure}

Fig. \ref{channel_capactiy_active_ports} illustrates the channel capacity results concerning the physical properties of the FAS. When the port spacing is held constant, the channel capacity is positively correlated with the proportion of active ports, intuitively demonstrating that a higher number of active ports provides more abundant transmission resources. In contrast, channel capacity is relatively insensitive to changes in port spacing. Even when the port spacing is adjusted, the resulting fluctuation in channel capacity is significantly smaller than the impact caused by variations in the active port ratio. Combined with the trends observed in Fig. \ref{fig4}, it is confirmed that the proportion of active ports plays a dominant role in determining the system performance of FAS-enabled UAV communications.

\begin{figure}[!t]
	\centering
	\includegraphics[width=\columnwidth]{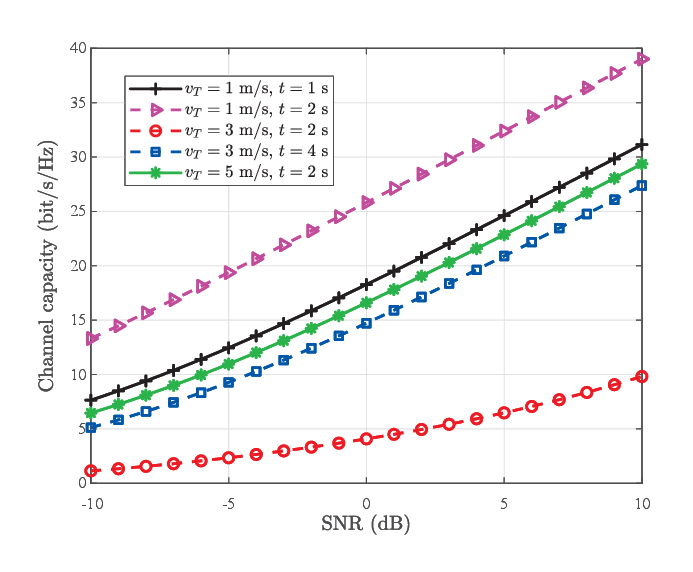}
	\caption{Channel capacity of the proposed FAS-assisted UAV-to-MU channel model with different motion state.}\label{channel_capactiy_motion}
	\vspace{-3mm}
\end{figure}

Fig. \ref{channel_capactiy_motion} provides the channel capacity results under various UAV motion states. The results reveal that the dynamic characteristics of the UAV, such as its velocity and flight duration, significantly influence the channel capacity behavior. Specifically, the channel capacity fluctuates with the UAV's flight speed, which is attributed to the combined effects of Doppler shifts and rapid variations in the channel environment. Furthermore, as the motion time increases, the continuous evolution of channel conditions consequently impacts the channel capacity. These findings indicate that the dynamic characteristics of the UAV must be rigorously accounted for in the design and deployment of FAS-enabled UAV communication systems.

\begin{figure}[!t]
	\centering
	\includegraphics[width=\columnwidth]{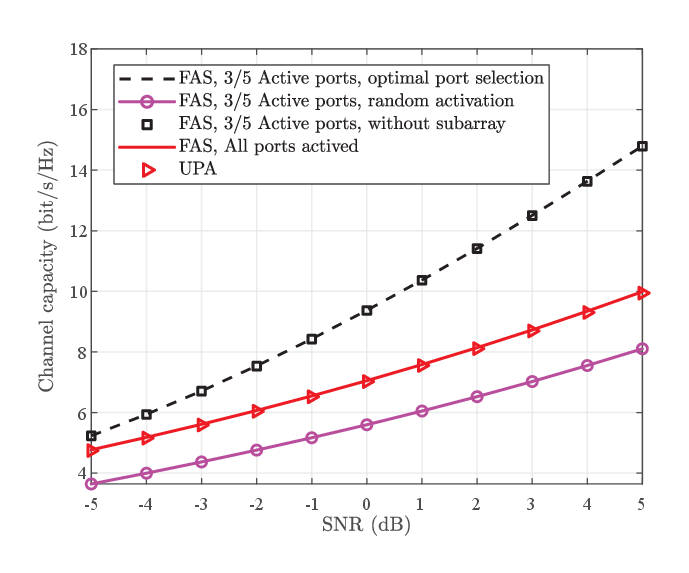}
	\caption{Channel capacity comparison under different FAS port activation and subarray partition schemes.}\label{channel_capactiy_subarray}
	\vspace{-3mm}
\end{figure}
Fig. \ref{channel_capactiy_subarray} compares the channel capacity of the proposed FAS with that of the UPA under different port activation strategies. For comparison, a random activation baseline with the same activation ratio of 3/5 is included. The proposed optimal port selection achieves higher channel capacity by prioritizing high-gain ports and avoiding ports in deep-fading regions.In addition, the capacity of the FAS with optimal port selection is higher than that of the UPA, demonstrating the performance advantage of FAS. It is also worth noting that when all FAS ports are activated, the channel capacity decreases compared with the optimal 3/5 activation case, since all-port activation includes ports with poor channel gains. Furthermore, the curves with and without the proposed subarray partition scheme are close to each other, indicating that the capacity degradation caused by the subarray partition is relatively limited in the considered scenario. These results verify that the performance gain of FAS comes not only from flexible port activation, but also from properly selecting high-gain ports according to the spatially non-uniform channel distribution.

\begin{figure}[!t]
	\centering
	\includegraphics[width=\columnwidth]{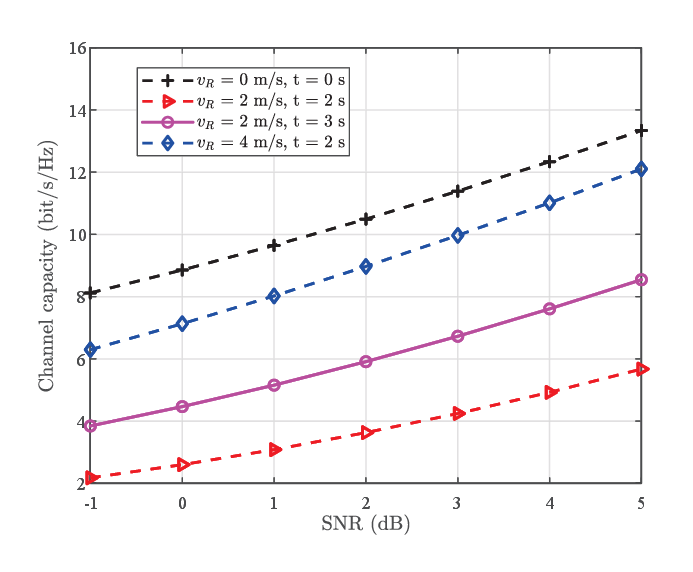}
	\caption{Impact of MU mobility on channel capacity.}\label{channel_capactiy_MU}
	\vspace{-3mm}
\end{figure}

Fig. \ref{channel_capactiy_MU} shows the channel capacity under different MU mobility conditions. The capacity increases with SNR in all cases, while different MU velocities and time instants lead to distinct capacity variations. This indicates that MU mobility changes the relative UAV-MU geometry and introduces additional temporal non-stationarity into the near-field channel. Therefore, although the baseline model assumes a low-mobility or quasi-static MU, high-mobility MU scenarios can be accommodated by using a time-varying MU position and performing more frequent CSI, port selection, and subarray updates.

\section{Conclusion}

In this paper, we proposed a dynamic port-reconfigurable channel model tailored for FAS-assisted UAV-to-MU near-field communications. Specifically, for the FAS mounted on the UAV, we designed a subarray partition scheme based on a greedy strategy, complemented by a dynamic update mechanism. Numerical results demonstrate that the proposed model accurately captures the non-uniform spatial distribution characteristics of near-field channels and exhibits robust adaptability to the non-stationary channel variations induced by UAV mobility. It was observed that FAS configuration parameters, such as port spacing and activation ratio, along with UAV dynamic characteristics, including velocity and flight duration, significantly influence system performance.
Specifically, the port spacing and active-port ratio significantly influence the spatial channel characteristics and the resulting channel capacity.
Similarly, maintaining the port activation ratio within an optimal range ensures spatial diversity gain while mitigating the performance degradation caused by deep fading ports. Furthermore, the efficacy of the greedy strategy-based subarray partition scheme was validated. This scheme significantly reduces computational complexity while guaranteeing the accuracy of near-field channel modeling. Finally, the accompanying dynamic update mechanism facilitates rapid adaptation to scenarios where new ports are activated, effectively eliminating the latency overhead associated with global re-partitioning.

As future work, we can point out two potential directions: (i) explore lightweight AI/hybrid learning based greedy partition for dynamic FAS-enabled UAV communication; and (ii) develop more advanced port selection methods by considering CSI acquisition overhead.

\appendix
\section{APPENDIX}
The channel matrix in \eqref{eq:H_all_ports} can be further expressed as
\begin{align}\label{eq:H_all_ports_c}
\mathbf{H} &(t,\tau) = \mathbf{H}^{\text{LoS}} (t,\tau) + \mathbf{H}^{\text{NLoS}} (t,\tau) \nonumber \\[0.125cm]
&= h^{\text{LoS}} (t) \mathbf{v}^{\text{T}}_{\text{LoS}} (t) \mathbf{u}^{\text{T}}_{\text{LoS}} (t) \delta(\tau - \tau^{\text{LoS}} (t)) \nonumber \\[0.125cm]
&+ \sum_{\ell \in L} \sum^{\ell_N}_{n = 1} h^{\text{NLoS}} (t) \mathbf{v}^{\text{T}}_{\text{NLoS}} (t) \mathbf{u}^{\text{T}}_{\text{LoS}} (t) \delta(\tau - \tau^{\text{NLoS}} (t)) \, .
\end{align}
where $h^{\text{LoS}} (t)$ and $h^{\text{NLoS}} (t)$ are the channel transmission parameters for the LoS and NLoS propagation links, respectively. $\mathbf{u}^{\text{T}}_{\text{LoS}} (t)$ and $\mathbf{v}^{\text{T}}_{\text{LoS}} (t)$ are the response vectors of the LoS propagation link at the UAV and the MU side, respectively. At the FAS side, To simplify channel-related calculations, we perform 1D vectorization on the original horizontal-vertical 2D port array of the FAS. Specifically, we first arrange all horizontal ports at the same vertical position in sequence, then process the ports at each vertical position in turn, and finally convert the 2D port matrix into a 1D vector, thus we have
\begin{align}
\mathbf{u}_{\text{LoS}} (t) = \big[ &e^{j \pi k_{1,1} \theta^{\text{azi}}_T (t) \theta^{\text{ver}}_T (t) } , ... , e^{j \pi k_{P_h,1} \theta^{\text{azi}}_T (t) \theta^{\text{ver}}_T (t) },\nonumber \\[0.125cm]
&e^{j \pi k_{1,2} \theta^{\text{azi}}_T (t) \theta^{\text{ver}}_T (t) } , ... , e^{j \pi k_{P_h,2} \theta^{\text{azi}}_T (t) \theta^{\text{ver}}_T (t) },\nonumber \\[0.125cm]
&..., \nonumber \\[0.125cm]
&e^{j \pi k_{1,P_v} \theta^{\text{azi}}_T (t) \theta^{\text{ver}}_T (t) } , ... , e^{j \pi k_{P_h,P_v} \theta^{\text{azi}}_T (t) \theta^{\text{ver}}_T (t) } \big] \, ,
\end{align}
where $\theta^{\text{azi}}_T (t)$ and $\theta^{\text{ver}}_T (t)$ represent the spatial frequencies along the azimuth and vertical dimensions with the path from the midpoint of the FAS at the UAV side to the ULA at the MU side, respectively. which can be expressed as
\begin{align}
\theta^{\text{azi}}_T (t) = \Delta d_h/\lambda (\cos{\alpha^{\text{LoS}}_T (t)} - \psi^{\text{azi}}_T) \cos{\beta^{\text{LoS}}_T (t) \cos{\psi^{\text{ver}}_T}} \, ,
\end{align}
\begin{align}
\theta^{\text{ver}}_T (t) = \Delta d_v/\lambda \sin{\beta^{\text{LoS}}_T (t) \sin{\psi^{\text{ver}}_T}} \, .
\end{align}

At the MU side, the channel transmission parameter can be expressed as
\begin{align}
\mathbf{v}_{\text{LoS}} (t) = \big[ &e^{j \pi k_1 \theta^{\text{azi}}_R (t) \theta^{\text{ver}}_R (t) } , e^{j \pi k_2 \theta^{\text{azi}}_R (t) \theta^{\text{azi}}_R (t) \theta^{\text{ver}}_R (t) },\nonumber \\[0.125cm]
&\hspace{2.6cm}..., e^{j \pi k_Q \theta^{\text{azi}}_R (t) \theta^{\text{ver}}_R (t) } \big] \, ,
\end{align}
where $\theta^{\text{azi}}_R (t)$ and $\theta^{\text{ver}}_R (t)$ represent the spatial frequencies along the azimuth and vertical dimensions with the path from the midpoint of the ULA at the MU side to FAS at the UAV side, respectively, which can be expressed as
\begin{align}
\theta^{\text{azi}}_R (t) = \delta_R/\lambda (\cos{\alpha^{\text{LoS}}_R (t)} - \psi^{\text{azi}}_R) \cos{\beta^{\text{LoS}}_R (t)} \cos{\psi^{\text{ver}}_R} \, ,
\end{align}
\begin{align}
\theta^{\text{ver}}_R (t) = \delta_R/\lambda \sin{\beta^{\text{LoS}}_R (t)} \sin{\psi^{\text{ver}}_R} \, .
\end{align}

For the NLoS propagation link, $\mathbf{u}^{\text{T}}_{\text{NLoS}} (t)$ and $\mathbf{v}^{\text{T}}_{\text{NLoS}} (t)$ denote the response vectors of at the UAV and the MU side, respectively, which can be expressed as
\begin{align}
\mathbf{u}&_{\text{NLoS}} (t) = \nonumber \\[0.125cm]
&\big[ e^{j \pi k_{1,1} \theta^{\text{azi}}_{T,\ell_n} (t) \theta^{\text{ver}}_{T,\ell_n} (t) } ,...,  e^{j \pi k_{P_h,1} \theta^{\text{azi}}_{T,\ell_n} (t) \theta^{\text{ver}}_{T,\ell_n} (t)},\nonumber \\[0.125cm]
&e^{j \pi k_{1,2} \theta^{\text{azi}}_{T,\ell_n} (t) \theta^{\text{ver}}_{T,\ell_n} (t) } ,...,  e^{j \pi k_{P_h,2} \theta^{\text{azi}}_{T,\ell_n} (t) \theta^{\text{ver}}_{T,\ell_n} (t)},\nonumber \\[0.125cm]
&...,\nonumber \\[0.125cm]
&e^{j \pi k_{1,P_v} \theta^{\text{azi}}_{T,\ell_n} (t) \theta^{\text{ver}}_{T,\ell_n} (t) } ,...,  e^{j \pi k_{P_h,P_v} \theta^{\text{azi}}_{T,\ell_n} (t) \theta^{\text{ver}}_{T,\ell_n} (t)} \big] \, ,
\end{align}
where $\theta^{\text{azi}}_{T,\ell_n} (t)$ and $\theta^{\text{ver}}_{T,\ell_n} (t)$ represent the spatial frequencies along the azimuth and vertical dimensions with the path from the midpoint of the FAS at the UAV side to the $n$-th path in the $\ell$-th cluster, respectively, which can be expressed as
\begin{align}
\theta^{\text{azi}}&_{T,\ell_n} (t) = \nonumber \\[0.125cm]
&\Delta d_h/\lambda (\cos{\alpha^{\text{NLoS}}_{T,\ell_n} (t)} - \psi^{\text{azi}}_T) \cos{\beta^{\text{NLoS}}_{T,\ell_n} (t)} \cos{\psi^{\text{ver}}_T} \, ,
\end{align}
\begin{align}
\theta^{\text{ver}}_{T,\ell_n} (t) = \Delta d_v/\lambda \sin{\beta^{\text{NLoS}}_{T,\ell_n} (t)} \sin{\psi^{\text{ver}}_T} \, .
\end{align}

At the MU side, the channel transmission parameter can be expressed as
\begin{align}
&\mathbf{v}_{\text{NLoS}} (t) = \nonumber \\[0.125cm]
&\big[ e^{j \pi k_1 \theta^{\text{azi}}_{R,\ell_n} \theta^{\text{ver}}_{R,\ell_n} } , e^{j \pi k_2 \theta^{\text{azi}}_{R,\ell_n} \theta^{\text{ver}}_{R,\ell_n} },..., e^{j \pi k_Q \theta^{\text{azi}}_{R,\ell_n} \theta^{\text{ver}}_{R,\ell_n} } \big] \, ,
\end{align}
where $\theta^{\text{azi}}_{R,\ell_n} (t)$ and $\theta^{\text{ver}}_{R,\ell_n} (t)$ represent the spatial frequencies along the azimuth and vertical dimensions with the path from the midpoint of the ULA at the MU side to FAS at the UAV side, respectively, which can be expressed as
\begin{align}
\theta^{\text{azi}}_{R,\ell_n} = \delta_R/\lambda (\cos{\alpha^{\text{NLoS}}_{R,\ell_n}} - \psi^{\text{azi}}_R) \cos{\beta^{\text{NLoS}}_{R,\ell_n}} \cos{\psi^{\text{ver}}_R} \, ,
\end{align}
\begin{align}
\theta^{\text{ver}}_{R,\ell_n} = \delta_R/\lambda \sin{\beta^{\text{NLoS}}_{R,\ell_n}} \sin{\psi^{\text{ver}}_R} \, .
\end{align}

\bibliographystyle{IEEEtran}

\begin{thebibliography}{10}
\providecommand{\url}[1]{#1}
\csname url@samestyle\endcsname
\providecommand{\newblock}{\relax}
\providecommand{\bibinfo}[2]{#2}
\providecommand{\BIBentrySTDinterwordspacing}{\spaceskip=0pt\relax}
\providecommand{\BIBentryALTinterwordstretchfactor}{4}
\providecommand{\BIBentryALTinterwordspacing}{\spaceskip=\fontdimen2\font plus
\BIBentryALTinterwordstretchfactor\fontdimen3\font minus
  \fontdimen4\font\relax}
\providecommand{\BIBforeignlanguage}[2]{{%
\expandafter\ifx\csname l@#1\endcsname\relax
\typeout{** WARNING: IEEEtran.bst: No hyphenation pattern has been}%
\typeout{** loaded for the language `#1'. Using the pattern for}%
\typeout{** the default language instead.}%
\else
\language=\csname l@#1\endcsname
\fi
#2}}
\providecommand{\BIBdecl}{\relax}
\BIBdecl


\bibitem{Saad-2020}
W. Saad, {\em et al.}, ``A vision of 6G wireless systems: Applications, trends, technologies, and open research problems,'' {\em IEEE Netw.}, vol. 34, no. 3, pp. 134-142, May 2020.

\bibitem{Wang-2023}
C.-X. Wang, {\em et al.}, ``On the road to 6G: Visions, requirements, key technologies, and testbeds,'' {\em IEEE Commun. Surv. Tutorials}, vol. 25, no. 2, pp. 905-974, Secondquarter 2023.

\bibitem{Liu-2021}
C. Liu, {\em et al.}, ``Cell-free satellite-UAV networks for 6G wide-area internet of things,'' {\em IEEE J. Sel. Areas Commun.}, vol. 39, no. 4, pp. 1116-1131, Apr. 2021.

\bibitem{Aggarwal-2021}
S. Aggarwal, {\em et al.}, ``Blockchain-envisioned UAV communication using 6G networks: Open issues, use cases, and future directions,'' {\em IEEE Internet Things J.}, vol. 8, no. 7, pp. 5416-5441, Apr. 2021.

\bibitem{Jiang-2025DL}
H. Jiang, {\em et al.}, ``Deep reinforcement learning enabled UAV trajectory optimization for A2G communication systems,'' {\em IEEE Trans. Cogn. Commun. Netw.}, vol. 12, pp. 3164-3178, Nov. 2026.

\bibitem{Wang-2024MIMO}
H. Wang, {\em et al.}, ``MIMO channel spatial correlation and capacity in tunnel entrance scenarios,'' {\em IEEE Antennas Wireless Propag. Lett.}, vol. 23, no. 1, pp. 319-323, Jan. 2024.

\bibitem{Wang-2024}
Z. Wang, {\em et al.}, ``Extremely large-scale MIMO: Fundamentals, challenges, solutions, and future directions,'' {\em IEEE Wireless Commun.}, vol. 31, no. 3, pp. 117-124, Jun. 2024.

\bibitem{ZhangMIMO}
Z. Zhang \emph{et al.}, ``Efficient ODMA for unsourced random access in MIMO and hybrid massive MIMO," \emph{IEEE Internet Things J.}, vol. 11, no. 23, pp. 38846-38860, 1 Dec.1, 2024.
\bibitem{ZhangMIMO2}
Z. Zhang, \emph{et al.}, ``Joint pattern, data and channel estimation for unsourced random access in GMAC and MIMO systems,'' \emph{IEEE Trans. Wireless Commun.}, Early Access, DOI: 10.1109/TWC.2026.3705749.

\bibitem{Wong-2021}
K. K. Wong, {\em et al.}, ``Fluid antenna systems,'' {\em IEEE Trans. Wireless Commun.}, vol. 20, no. 3, pp. 1950-1962, Mar. 2021.

\bibitem{Jiang-2025FAS}
H. Jiang, {\em et al.}, ``Dynamic channel modeling of fluid antenna systems in UAV communications,'' {\em IEEE Wireless Commun. Lett.}, vol. 14, no. 10, pp. 3169-3173, Oct. 2025.

\bibitem{Zhang-2025JSAC}
Z. Zhang, {\em et al.}, ``On fundamental limits for fluid antenna-assisted integrated sensing and communications for unsourced random access,'' {\em IEEE J. Sel. Areas Commun.}, vol. 44, pp. 136-149, Sep. 2026.

\bibitem{Ghadi-2025}
F. R. Ghadi, {\em et al.}, ``UAV-relay assisted RSMA fluid antenna system: Outage probability analysis,'' {\em IEEE Wireless Commun. Lett.}, vol. 14, no. 9, pp. 2907-2911, Sept. 2025.

\bibitem{Wong2020}
K. K. Wong, {\em et al.}, ``Fluid antenna system for 6G: When Bruce Lee inspires wireless communications,'' {\em IET Electron. Lett.}, vol. 56, no. 24, pp. 1288-1290, Nov. 2020.

\bibitem{Wong-2020fas}
K. K. Wong, {\em et al.}, ``Performance limits of fluid antenna systems,'' {\em IEEE Commun. Lett.}, vol. 24, no. 11, pp. 2469-2472, Nov. 2020.

\bibitem{Zhang-2026}
Z. Zhang, {\em et al.}, ``Finite-aperture fluid antenna array design: Analysis and algorithm,'' {\em IEEE IEEE Wireless Commun. Lett.}, vol. 15, pp. 3199-3203, 2026.
\bibitem{FAA2}
 Z. Zhang, {\em et al.}, ``Finite-aperture planar fluid antenna array,'' \emph{preprint, arXiv:2605.22040}, 2026.
\bibitem{FAA3}
Z. Zhang, {\em et al.}, ``Electromagnetic-aware fluid antenna array,'' \emph{preprint, arXiv:2607.21375}, 2026.
\bibitem{CKJ1}
K. Chen, {\em et al.}, ``DBRAA: Sub-6 GHz and millimeter wave dual-band reconfigurable antenna array for ISAC,'' \emph{IEEE Trans. Commun.}, vol. 73, no. 10, pp. 9830-9845, Oct. 2025.
\bibitem{CKJ2}
K. Chen, {\em et al.}, ``Beam training and tracking for extremely large-scale MIMO communications,'' \emph{IEEE Trans. Wireless Commun.}, vol. 23, no. 5, pp. 5048-5062, May 2024.
\bibitem{CKJ3}
K. Chen, C. Qi, O. A. Dobre, and G. Y. Li, ``Simultaneous beam training and target sensing in ISAC systems with RIS,'' \emph{IEEE Trans. Wireless Commun.}, vol. 23, no. 4, pp. 2696-2710, Apr. 2024.

\bibitem{Zhang-2025}
Z. Zhang, {\em et al.}, ``On fundamental limits of slow-fluid antenna multiple access for unsourced random access,'' {\em IEEE Wireless Commun. Lett.}, vol. 14, no. 11, pp. 3455-3459, Nov. 2025.
\bibitem{AD_z}
Z. Zhang, \emph{et al.}, ``Cramer-Rao Bounds for Activity Detection in Conventional and Fluid Antenna Systems'', \emph{IEEE Wireless Commun. Lett.}, {\em IEEE Wireless Commun. Lett.}, vol. 15, pp. 3059-3063, 2026.
\bibitem{Wong-2022}
K. K. Wong and K. F. Tong, ``Fluid antenna multiple access,'' {\em IEEE Trans. Wireless Commun.}, vol. 21, no. 7, pp. 4801-4815, July 2022.

\bibitem{Alvim2023on}
P. D. Alvim, {\em et al.}, ``On the performance of fluid antennas systems under $\alpha$-$\mu$ fading channels,'' {\em IEEE Wireless Commun. Lett.}, vol. 13, no. 1, pp. 108-112, Jan. 2024.

\bibitem{New2023fluid}
W. K. New, {\em et al.}, ``Fluid antenna system: New insights on outage probability and diversity gain,'' {\em IEEE Trans. Wireless Commun.}, vol. 23, no. 1, pp. 128-140, Jan. 2024.

\bibitem{Zhang-2025CSI}
Z. Zhang, {\em et al.}, ``Joint activity detection and channel estimation for fluid antenna system exploiting geographical and angular information,'' \emph{IEEE J. Sel. Topics Signal Process.}, vol. 20, no. 3, pp. 354--370, 2026.
\bibitem{Zhang-2025CSI2}
Z. Zhang, {\em et al.}, ``Geometry-structured channel reconstruction for conventional and fluid antenna systems: Bayesian inference and fundamental limits,'' \emph{arXiv preprint arXiv:2606.04001}, 2026.

\bibitem{Shen-2025}
L.-H. Shen and Y.-H. Chiu, ``RIS-aided fluid antenna array-mounted AAV networks,'' {\em IEEE Wireless Commun. Lett.}, vol. 14, no. 4, pp. 1049-1053, Apr. 2025.

\bibitem{Abdou-2024}
S. B. S. Abdou, {\em et al.}, ``Sum-rate maximization for UAV relay-aided fluid antenna system with NOMA,'' in {\em Proc. IEEE Int. Symp. Telecommun. Technol. (ISTT)}, Langkawi Island, Malaysia, 2024, pp. 53-58.


\bibitem{Han-2024}
C. Han, {\em et al.}, ``Cross far- and near-field wireless communications in terahertz ultra-large antenna array systems,'' {\em IEEE Wireless Commun.}, vol. 31, no. 3, pp. 148-154, Jun. 2024.

\bibitem{Cui-2024}
M. Cui and L. Dai, ``Near-field wideband beamforming for extremely large antenna arrays,'' {\em IEEE Trans. Wireless Commun.}, vol. 23, no. 10, pp. 13110-13124, Oct. 2024.

\bibitem{Liu-2023}
Y. Liu, {\em et al.}, ``Near-field communications: A tutorial review,'' {\em IEEE Open J. Commun. Soc.}, vol. 4, pp. 1999-2049, Aug. 2023.

\bibitem{Wan-2024}
Z. Wan, {\em et al.}, ``Near-field channel modeling for electromagnetic information theory,'' {\em IEEE Trans. Wireless Commun.}, vol. 23, no. 12, pp. 18004-18018, Dec. 2024.

\bibitem{Ruan-2024}
C. Ruan, {\em et al.}, ``Wideband near-field channel covariance estimation for XL-MIMO systems in the face of beam split,'' {\em IEEE Trans. Veh. Technol.}, vol. 74, no. 2, pp. 2912-2926, Feb. 2025.

\bibitem{Yang-2024}
Y. Yang, {\em et al.}, ``Characteristics and channel capacity studies of a novel 6G non-stationary massive MIMO channel model considering mutual coupling,'' {\em IEEE J. Sel. Areas Commun.}, vol. 42, no. 6, pp. 1519-1533, Jun. 2024.

\bibitem{Jiang-2025TWC}
H. Jiang, {\em et al.}, ``Large-scale RIS enabled air-ground channels: Near-field modeling and analysis,'' {\em IEEE Trans. Wireless Commun.}, vol. 24, no. 2, pp. 1074-1088, Feb. 2025.

\bibitem{Jiang-2025IoT}
H. Jiang, {\em et al.}, ``High-efficient near-field channel characteristics analysis for large-scale MIMO communication systems,'' {\em IEEE Internet Things J.}, vol. 12, no. 6, pp. 7446-7458, Mar. 2025.

\bibitem{Peng-2022}
K. Peng, {\em et al.}, ``Reliability-aware computation offloading for delay-sensitive applications in MEC-enabled aerial computing,'' {\em IEEE Trans. Green Commun. Networking}, vol. 6, no. 3, pp. 1511-1519, Sept. 2022.

\bibitem{Chen-2026}
J. Chen {\em et al.}, ``Generative AI-aided QoE-aware resource allocations for RlS-assisted digital twin interaction with uncertain evolution,'' {\em IEEE Trans. Mob. Comput.}, vol. 25, no. 6, pp. 7888-7905, June 2026.

\bibitem{Zhou-2026}
L. Zhou, {\em et al.}, ``Digital twins for low-altitude UAV networks–cooperation and learning,'' {\em IEEE Trans. Mob. Comput.}, vol. 25, no. 4, pp. 4839-4856, Apr. 2026.


\bibitem{Espinosa2024}
P. Ram\'{i}rez-Espinosa, {\em et al.}, ``A new spatial block-correlation model for fluid antenna systems,'' {\em IEEE Trans. Wireless Commun.}, vol. 23, no. 11, pp. 15829-15843, Nov. 2024.

\bibitem{new2023information}
W. K. New, {\em et al.}, ``An information-theoretic characterization of MIMO-FAS: Optimization, diversity-multiplexing tradeoff and $q$-outage capacity,'' {\em IEEE Trans. Wireless Commun.}, vol. 23, no. 6, pp. 5541-5556, Jun. 2024.

\bibitem{Liu2025FAS}
B. Liu, {\em et al.}, ``Meta fluid antenna: Architecture design, performance analysis, experimental examination'', available: arXiv:2509.12032, 2025.

\bibitem{Zhang-2025opt}
Z. Zhang, {\em et al.}, ``Finite-blocklength fluid antenna systems with spatial block-correlation channel model,'' {\em IEEE Commun. Lett.}, vol. 15, pp. 1911-1915, Feb. 2026.

\bibitem{New-2025opt}
W. K. New, {\em et al.}, ``Channel estimation and reconstruction in fluid antenna system: Oversampling is essential,'' {\em IEEE Trans. Wireless Commun.}, vol. 24, no. 1, pp. 309-322, Jan. 2025.

\bibitem{Xu2024}
H. Xu, {\em et al.}, ``Channel estimation for FAS-assisted multi-user mmWave systems,'' {\em IEEE Commun. Lett.}, vol. 28, no. 3, pp. 632-636, Mar. 2024.








\end{thebibliography}
\balance


\end{document}